\documentclass{bmcart}

\usepackage[utf8]{inputenc} 
\usepackage{graphicx}
\usepackage{booktabs} 
\usepackage{caption}
\usepackage{subcaption}
\usepackage{url}

\startlocaldefs
\endlocaldefs

\begin{document}

\begin{frontmatter}

\begin{fmbox}
\dochead{Research}


\title{Understanding News Outlets' Audience-Targeting Patterns}


\author[
   addressref={aff1},                   
   corref={aff1},                       
   email={eelejalde@udec.cl}   
]{\inits{EE}\fnm{Erick} \snm{Elejalde}}
\author[
   addressref={aff2,aff3},
   email={lferresh@udd.cl}
]{\inits{LF}\fnm{Leo} \snm{Ferres}}
\author[
   addressref={aff4},
   email={schifane@di.unito.it}
]{\inits{RS}\fnm{Rossano} \snm{Schifanella}}


\address[id=aff1]{
  \orgname{Department of Computer Science, University of Concepcion}, 
  \city{Concecpci\'on},                              
  \cny{Chile}                                    
}
\address[id=aff2]{%
  \orgname{Data Science Institute, Universidad del Desarrollo},
  \city{Santiago},
  \cny{Chile}
}
\address[id=aff3]{%
  \orgname{Telef\'onica R\&D},
  \city{Santiago},
  \cny{Chile}
}

\address[id=aff4]{%
  \orgname{University of Turin},
  \city{Turin},
  \cny{Italy}
}


\begin{artnotes}
\end{artnotes}

\end{fmbox}


\begin{abstractbox}

\begin{abstract} 
The power of the press to shape the informational landscape of a population is unparalleled, even now in the era of democratic access to all information outlets. However, it is known that news outlets (particularly more traditional ones) tend to discriminate who they want to reach, and who to leave aside. In this work, we attempt to shed some light on the audience targeting patterns of newspapers, using the Chilean media ecosystem. First, we use the gravity model to analyze geography as a factor in explaining audience reachability. This shows that some newspapers are indeed driven by geographical factors (mostly local news outlets) but some others are not (national-distribution outlets). For those which are not, we use a regression model to study the influence of socioeconomic and political characteristics in news outlets adoption. We conclude that indeed larger, national-distribution news outlets target populations based on these factors, rather than on geography or immediacy.
\end{abstract}




\end{abstractbox}
%

\end{frontmatter}



\section{Introduction}


The mass media is one of the social forces with the strongest transformative power. However, news reach people unequally, in part because news outlets target content to certain audiences. According to Herman and Chomsky's Propaganda Model (PM)~\cite{herman1988manufacturing}, there are many factors that shape the distribution and influence of news media coverage. Two of the most important factors are the geographic reach of newspapers (national versus regional newspapers), and the direct targeting of specific sectors of the population and/or the political ideology of the media outlet itself. The PM states that each linguistic account of an event must pass through five filters that define what is newsworthy. One such filter, the {\em advertising} filter (number two in the PM) predicts that most news outlets will try to reach a specific audience (segments of the population) with the objective of maximizing profit, instead of actually informing. In other words, outlets will try to cater to the target demographic's expectations, rather than being fair in their treatment of what is news. For instance, some advertisers will prefer to take their businesses to outlets with target audiences of high purchasing power, which will eventually marginalize working-class audiences; or by political reasons, with advertisers declining to do business with outlets perceived as ideological enemies or indeed any media unfavorable to their private interests. 

Prat and Str{\"o}mberg~\cite{prat2011political} provide another model (henceforth PS) that seems to better define the same concept of a media system entirely driven by profit-maximization (PM's second filter). The authors identify the characteristics that a certain audience needs to have for an outlet to create content that is relevant \textit{to that audience} (\textit{Proposition 4} in~\cite{prat2011political}). To make \textit{Proposition 4} more concrete, Prat and Str\"omberg~\cite{prat2011political} suggest that the main factors that influence the mass media coverage of an event are: (a) if the matter is of interest for a large group of people (a group may be characterized by a political stand, geographic location, ethnicity, etc.), (b) if it has a significant advertising potential (e.g., it may attract readers with a higher purchasing power), (c) if it is newsworthy to a group within easy reach (i.e., it is cheap to distribute news to that group). Thus, in a media system driven by profit, areas of low population density, minorities, and low-income classes will be relatively under-served and underrepresented in mainstream news coverage. This creates a loop that ends up neglecting (policy-wise) the most vulnerable segments of society just for having limited access to the media.

To date, there has been comparatively little large-scale, quantitative research on the relationship between the quality and diversity of the contents the media generate, and the socioeconomic indices of a particular area of coverage. In this paper, we try to find whether or not an outlet's coverage deviates from the purely geographic influence to a more sophisticated behavior involving the weight of political and socioeconomic interests for example, as operationalized by Prat and Str\"omberg. We examine the degree to which different geographic locations in the same country are covered by existing news outlets using Chilean social media data. We quantify how much of this coverage can be explained by a natural geographic targeting (e.g., local newspapers will give more importance to local news), and how much can be attributed to the politic/socioeconomic profile of the areas they serve. To find these coverage effects as predicted by the PM and PS models, we look for empirical evidence in the massive adoption of social networks. More specifically, we use statistical models that show how much of the distribution of Twitter followers can be explained based on the geographic, political and socioeconomic features of the different areas.

\section{Related Work}

In this section we give a short account of these targeting strategies: geographic, socioeconomic, and political. We use the concepts in the PM, and the operationalization of PS to identify them.

\subsection{Geographic targeting}

According to Zipf's \textit{Gravity Model}~\cite{10.2307/1417611,10.2307/2087063}, as we move farther away from the source of a piece of news, the interest/relevance of a story drops. Given that news outlets tend to cover stories where reporters can get quickly and easily (again, to minimize the cost of the piece of news), their followers are expected to be predominantly from populations that are closer to them. Also, the size of a population at a particular place may influence how newspapers cover events originating in that area. Newspapers work on an economy of scale with a considerable first copy cost. According to the Gravity Model, we could predict the flow of information in the news media system and hence, indirectly, the proportional distribution of followers a target area will have for a given news outlet.

Distance and population size are also essential magnitudes to describe profit in the model proposed by Prat and Str\"omberg (\textit{Equation 5} in~\cite{prat2011political}). News outlets will favor in their coverage issues that may draw the attention of larger groups (e.g., big cities) and to which it is cheaper to deliver the news (e.g., at a shorter distance).

\subsection{Socioeconomic targeting}
Another factor that influences the news coverage is the socioeconomic profile of an area. As we mentioned earlier, a strategy that the news outlets could implement to increase advertising revenue is to target sectors of the population with a higher purchasing power. Herman and Chomsky point out in the second filter of the PM that advertising, being a fundamental source of income for news outlets, plays an important role to maintain the hegemony of the top news companies in the free market. News outlets that can secure good advertising contracts may afford lower sell prices and become more competitive. This business model breaks the natural market rules that give the final buyer's choice the power to decide. In this case, the advertisers' contracts have a significant impact on the media growth or even their survival. So, outlets are forced to comply and demonstrate to the announcer how their content may serve to its needs. The audience of a newspaper becomes its product, which can then be ``sold'' to the sponsors.

The second filter of the PM is in line with the predictions in the Proposition 4(b) in~\cite{prat2011political}. This filter suggests that in their effort to align their content with the advertisers' interests, the media have shifted to a lighter and less controversial programming (e.g., lifestyle, fashion, sports, etc.)~\cite{news.coverage_news.interest_a,news.coverage_news.interest_b}. In~\cite{10.2307/j.ctt7smgs}, the author presents some evidence on the same direction, showing, for example, media preference for ``soft news'' content that is favored by advertisers as it targets a demographic of female and young people. A more recent and direct example on how advertisers may influence the content of the media is the evolution from product placements to \textit{Native Ads}~\cite{wiki.native.ads}, which makes it difficult to the reader to differentiate between news and advertisement. This type of pseudo-content provides a significant part of the outlets' revenue~\cite{forbes:native-ads}.

Being able to detect this kind of behavior in the media is of utmost importance. For example, a socioeconomic bias in the media system can be very damaging as it may exacerbate the gap between rich and poor areas. A population with limited access to the news is less informed and, consequently, less likely to hold authorities responsible for public expenditure and providing broad public welfare~\cite{doi:10.1162/jeea.2005.3.2-3.259,doi:10.1093/wbro/lki002}. In turn, this motivates the incumbent to prioritize and divert resources to places where they will receive more media coverage and not necessarily where they are most needed. According to Chomsky and Herman~\cite{herman1988manufacturing}, these characteristics make the news media system comparable to a political scheme where votes are weighted by income.

Other aspects of a community and their links to different socioeconomic conditions have been studied. For example, the diversity in the individuals' relationships~\cite{Eagle1029} or patterns in the urban mobility~\cite{Smith:2013:FPI:2441776.2441852} have shown to be useful indicators for the deprivation levels of a region. However, there has been comparatively little large-scale quantitative research on the relationship between the media coverage, and the socioeconomic indices of a particular area.

\subsection{Political targeting}
Political bias is probably the most studied type of bias in the mass media~\cite{Park:2012:CFM:2209310.2209311, ECTA:ECTA994, ICWSM112782, Sudhahar:2012:EDP:2380921.2380938}. In previous work~\cite{Elejalde2018}, we analyze the nature of bias through a political quiz. Our study shows that even the political bias could have some economic factors.
Extra evidence of this is given in \cite{ECTA:ECTA994}. The authors estimate the bias in newspapers according to how similar their language is compared to that used by congressmen for which a right/left stand is known. They do not find a direct relationship between the ``slant'' of a newspaper and the political preference of the owners (cf. our own work on the topic \cite{2017arXiv171006347B}). Instead, bias in the news is found to be more correlated to the political inclinations of the readers, showing a tendency in these news outlets to align themselves with the political preferences of their target audience and hence, maximizing selling profits. We think this is an important result because, although outlets may seem to take a political stand in their editorial line, evidence suggests that this may be another strategy to generate revenue by targeting a specific group of people. For example, governmental offices at various levels assign a considerable part of their budgets to advertising. Newspapers sympathizers of the government policies may benefit from lucrative advertising contracts with the incumbent. So, outlets discrimination can be also influenced by political reasons, with advertisers declining to do business with media that are perceived as ideological enemies or generally unfavorable to their interests.

\subsection{Online news dynamics, coverage, bias}
In an interview with Mullen~\cite{Mullen2009}, Herman and Chomsky express their confidence in the applicability of the PM to forms of media other than newspapers, especially the Internet, where traditional news outlets compete with new digital media and advertising is more relevant than ever before. If anything, the ``old'' news industry has evolved and has adapted to the new environment.

More recently, Robinson~\cite{Robinson2015} evaluates if the introduction of new communication technologies and the Internet have affected the influence of the economic structure over the news media system. He argues that even though there has been a shift from the printed news to the digital media (newspaper website audiences grew by 7.4\% in 2012~\cite{StateMedia2012}), the news cycle is still controlled by the big news corporations. The three most significant outlets in U.S. (i.e., Wall Street Journal, USA Today and New York Times) had a circulation of over 5 million users (which include  digital subscribers); any of them counting at least one order of magnitude more readers than the next closest competitor~\cite{StateMedia2012}. With regards to advertising on Internet, the author concludes that news outlets had to rely even harder on this source of income: on-line subscription revenue does not cover the previous earnings made from selling print newspapers. So, the content has became more profit driven with a shift to soft-news and corporate-friendly reports~\cite{StateMedia2012}.

On the same note, Pew Research Center reported that nowadays up to 93\% of U.S. adults consume some news online~\cite{StateMedia2017a}, and to do so they use official news organizations websites and social media in equal shares (36\% vs. 35\% respectively)~\cite{StateMedia2017b}. However, regardless of the path they use to get to news, readers overwhelmingly favor professional news organizations to get their news (76\%)~\cite{StateMedia2017d}. With the increasingly important role of social media in the news system, news outlets have again adapted by embracing the new platform and now they are able to create original content for them (i.e., Facebook's instant articles\footnote{https://instantarticles.fb.com/} allows any publisher to create a piece of news directly in the Facebook platform). This move is also deeply motivated by the market and the task to increase advertising revenue. In 2016 digital advertising constituted approximatively 37\% of all advertising revenue (not just news outlets), and Facebook along comprises for 35\% of digital advertising revenue~\cite{StateMedia2017a}.

More specifically, in~\cite{doi:10.1080/21670811.2015.1133249}, the authors analyze the dynamics of news and journalism on the Twitter platform. They found that $0.8\%$ of the tweets are news media related. This gives an idea of how significant are news media to Twitter. They also report, confirming the results of Robinson~\cite{Robinson2015}, that the traditional notion of gatekeeping and news production have not changed, and large news organizations still control what is newsworthy. Moreover, they show that news entities do not use the social media to engage with their audience but rather as a way for content dissemination (mostly by redirecting users to their own websites). Also, they find difference in interest (topic-wise) between Twitter users and news outlets. This difference of focus gives some evidence to the agenda-setting behavior of the news industry, reinforcing the hypothesis of a profit-driven system, instead of an informative one.

\section{Data}
\label{sec:geocov_data}

To create our database of news outlets, we used different sources, with Poderopedia's ``influence'' database~\cite{poderopedia} and Wikepedia~\cite{wiki.chilean.newspapers} as our baseline, manually adding other news outlets in Chile. Our database contains 403 \textit{active} accounts. An account is considered \textit{active} if it tweets at least once a month. We enriched the profile of each outlet by adding relevant information such as geographic location, scope, Twitter account, and number of Twitter followers.

The Twitter's API allows to automatically access the flow of tweets and query the system for user profiles, followers and tweeting history. This data availability makes it possible to explore the behavior and interactions of personal and institutional accounts, developing and testing social theories at a scale that was unfeasible few years ago. This is the closest thing we have to a record of the every-day life of over 300 million people (Twitter reported 328 million monthly active users in the first quarter of 2017~\cite{forbes:twitter}).

Chile ranks among the top-10 countries regarding the average number of Twitter users per 1000 individuals~\cite{10.1371/journal.pone.0061981}. Because of the massive adoption of Twitter and the strong presence of the news media on the social networks, we use the Twitter followers as a proxy for the actual audience of a news outlet. For our analysis, we download the user's profile on Twitter for each follower of the outlets in our database. We collected the profile of 4,943,351 unique users. Each user may simultaneously follow more than one news outlet.

We decided to use the commune as our location unit given that this is the smaller political division in Chile, but at the same time, it is big enough to create both a statistical and popularly perceived socioeconomic profile at the population level.


We obtained the total population of each commune and other demographic indices from the National Institute of Statistics (INE)~\cite{ine.demo}. The demographic indices were already aggregated by commune.
%
We also needed information on the socioeconomic development of each zone. This kind of information is harder to obtain. The most reliable source is the national census. The problem is that censuses are very expensive and therefore are performed very infrequently (sometimes more than a decade apart - last completed valid census performed in Chile was in 2002)\footnote{There was another census in 2012, but it was methodologically flawed, with problems in coverage, and a supposed manipulation of some of the key indices \cite{pnud2013}.}. Instead, we use the data from the National Socio-Economic Characterization Survey (CASEN) from 2013\footnote{There is a CASEN survey from 2015 but the expansion factor for the communes is not complete (not even for the communes in Santiago)}~\cite{casen2013}. This study is conducted by the Ministry of Social Development in Chile. From the CASEN survey, we obtain the socioeconomic indicator at the level of a commune using the available expansion factor to calculate the weighted average income per household.

Our last dimension has to do with the political leaning of the communes. To measure the political tendencies of each geographical area, we use the results from the presidential election. The Chilean Electoral Service~\cite{servel} provides detailed information district-wise on the Chilean presidential elections since 1989 (that is, since Chile's return to democracy after the dictatorship of Augusto Pinochet).

To help understand the collected data, we represent in Figure \ref{fig:maps} the geographic distribution of each dimension for the most populated region of the country (i.e., Santiago). Note that most of the news outlets are located in the city center (Figure \ref{fig:z-dist}) and surrounded by very densely populated areas (Figure \ref{fig:population}). Also, Figure \ref{fig:z-income} shows a difference in the income level between communes from East to West (which coincides with the popular perception). Finally, more populated and urbanized areas show a right-leaning political predominance. While areas farther from the city center and mostly rural show the opposite tendency (see Figure \ref{fig:z-right-leaning}).

\section{Methods}

\subsection{Geographic targeting}
We first use the Gravity Model to identify how much of news coverage can be explained by the geographic factors of distance and population. For this we use the population and location of both the source of the medium and the target area. Equation \ref{eq:myGravityModel} represents this relation.

\begin{equation} \label{eq:myGravityModel}
F_i^j = \frac{P_i*P_j}{D_{ij}}
\end{equation}

Here $P_i$ is the population of the commune $i$, and $P_j$ is the population of the commune in which outlet $j$ is located. $D_{ij}$ represents the distance between the two communes. Then, $F_i^j$ should give us a value that represents the expected number of followers that outlet $j$ will have in the commune $i$.

We run the model for each news outlet so we can analyze the geographic targeting behaviour for different types of media.

For this study, we manually locate each news outlets in its source commune. The location may be determined by the intended audience if the name of the commune is in the name of the outlets (e.g., \textit{soyConcepcion} is assigned to \textit{Concepcion} city) or by the location of its headquarters.
At the intra-country level, big news media companies may have more than one headquarter, however in most cases they either work under a different name (with a more ``local'' name) or report directly to the central headquarters which ultimately define the editorial line. For example, \textit{Soy Concepcion} is owned by the \textit{El Mercurio} Group, which is also the group that owns one of the largest newspapers of the capital region (also called \textit{El Mercurio}).

Finally, for every pair of communes we use their estimated populations (obtained from the INE~\cite{ine.demo}) and GPS coordinates. We calculate the direct distance between them using the Haversine formula. We represent each outlet $j$ as a vector $F^j$. The elements of $F^j$ are the predicted proportion of followers in each commune for outlet $j$ obtained from the Gravity Model. We also create a vector $T^j$ for each outlet with the actual number of Twitter followers on each commune $i$ obtained from our ground-truth (see Section \ref{sec:users_geoloc_methodology} below). Using the two vectors that represents each outlet we calculate the Pearson product-moment correlation coefficients. This coefficient will give us, for each news outlet, an idea of how much of the distribution of readers can be attributed to the geographic dimension.

\subsubsection{Geolocation of the followers}
\label{sec:users_geoloc_methodology}
Regarding the location of the Twitter followers, there is an extensive body of work that focuses on geo-tagging Twitter users~\cite{Ozdikis:2017:SLE:3125062.3125098,Ajao:2015:SLI:2879214.2879221,7998610}. Most of this work can be divided into two groups according to their approach: content-based and network-based. Methods based on content can be further subdivided into those that use a gazetteer~\cite{Amitay:2004:WGW:1008992.1009040}, as in our case, to find direct references to geographic places and those based on \textit{Language Models} that try to learn a probabilistic text model~\cite{Serdyukov:2009:PFP:1571941.1572025}.  The performance of the former depends heavily on the quality of the used dictionary. The latter may achieve high precision for the geo-localization of users at a country level, or even within country regions or cities~\cite{Cheng:2010:YYT:1871437.1871535, Ryoo:2014:ITU:2567948.2579236}. However, to achieve a good performance at a finer grain classification, such as commune/neighborhood level, massive corpora of social media annotation is required~\cite{7998610}. On the other hand, the geo-localization of users based on their network (based on the assumption that users are more likely to interact with other users that are geographically closer to them) are more accurate at a finer level~\cite{Backstrom:2010:FMY:1772690.1772698, Kong:2014:SLS:2733004.2733060}. The problem is that crawling the connections of several million users and dealing with the corresponding graph is time consuming and computationally intensive.

In this paper we decided to test our hypothesis using only the users that we were able to geolocate based on their profile's \textit{location} field. We use these as a sample of the population. The cumulative number of followers per commune in our sample is highly correlated with the actual population distribution ($r(343)=.61, p<.01$). 
So, we will use this information to model news outlets coverage in our database.

To find the accounts that are following more than one outlet we use the identifier from each user's profile on Twitter. We use only those profiles that have a non-empty {\it location} field, which brings our list down to 1,579,068 accounts (31\% of the initial amount).

In \cite{Hecht:2011:TJB:1978942.1978976} the authors analyze the nature of the {\it location} field in the Twitter profile. Given that this is an open text field, users not always enter a valid (or even geographic) information. So, a pre-processing of these data is in order if we are making any study involving geolocation of the users based on this field.

In our remaining 31\%, some of the users have GPS coordinates, and others have a text description of their location. Since the text description is a free text entered by the user, it ranges from an exact postal address to a completely useless text (e.g., ``The milky way''). Using a gazetteer, we could extract 996,326 users with a recognizable location, which represents the 20\% of the initial amount. We tried to assign each user to a commune with a given level of confidence. For the users with a pair of GPS coordinates, we used a shape-file~\cite{rulamahue} of the communes of Chile to find the one that enclosed the point. Only 4,829 of the users had GPS coordinates. The users with a text description making explicit mention of a commune were assigned to that commune. For those who mentioned only a province or a region, we could allocate them in the city/commune capital of that region. Given that these cities have the most prominent population density in the area, we would maximize the chance to be correct when making a guess.
Nevertheless, we choose to work only with users for which we have high confidence in their location, namely: those with GPS coordinates or explicit mention of a commune. Thus, our final list contains 602,810 users, which is over 12\% of the total number of unique followers (Table \ref{table:followers} summarizes the followers).

\subsection{Audience-targeting model}
\label{sec:pol-eco_methodology}
According to the PM and PS models, direct targeting of specific sectors of the population shape the distribution of news. If motivated by a profit-driven model of the media system, this targeting may be based in socioeconomic and/or political characteristics of the intended audience.

We use a regression model to study the influence of the different features that represent the dimensions in our hypothesis, namely the socioeconomic and political characteristics of the communes that may attract profit-driven media coverage. We try to model the ranking of communes for each outlet based on the share of followers from each commune. We include the geographic factor in our model to measure its influence and to keep a reference. We use as the geographic feature only the distance from the commune to the news source, given that the actual population is to closely related to our target variable (a function on the number of followers).
We estimate the socioeconomic level of an area as its expected household income. Meanwhile, for the political factor, we use the right/left-leaning of the commune (see section \ref{sec:geocov_data}).

To calculate the political factor per commune, we first aggregated the raw number of votes receives by each party on each commune in the past three elections. We manually annotated each political party as left-wing, right-wing or centrist according to their self-declared position. Political parallelism on the media system is seen when the media outlets are popularly perceived as leaned to one broad side in the political spectrum (not necessarily linked to a political party but rather to a political range)~\cite{hallin2004comparing}. So, we aggregated the votes for all parties that have a similar political ideology. With this, we measure how "left-leaning" or "right-leaning" is a commune.

The data in all three dimensions was aggregated at the level of communes and normalized by calculating the $z-score$ of each area on each feature.


For the model we use a random forest regressor~\cite{Breiman:2001:RF:570181.570182} (implemented in the module {\tt RandomForestRegressor} within the python library {\tt scikit-learn}). This estimator is based on classifying decision trees. Models based on decision trees are less susceptible to overfitting, considering that our training sets are relatively small (for each newspaper we only have as many samples as communes with a valid entry). 

We evaluate the model using a random shuffle cross validation that leaves 20\% of the dataset for testing, and trains the regressor in the remaining 80\%. Each experiment is repeated 100 times and the average score and standard deviation are reported. We measure the quality of the fit with the explained variance.

We also measure the explanatory power of each individual dimension on the media coverage. For this, we calculate the Kendall-Tau correlation of the corresponding feature against the number of followers per commune for each news outlet. The results of these measurements should give information on the marketing strategy of different outlets.

\subsubsection{Chilean national soccer team followers}
To validate the results and ensure that they are peculiar to the online media ecosystem and not an artifact of the social media attention dynamics as a whole, we repeat the same experiment using a different topical dataset. In particular, we gather data on the Twitter followers for the soccer players that were part of the Chilean national football\footnote{``Soccer'', in the US dialect.} team in the "Copa America Centenario 2016" tournament. We expect the coverage variable in a sport-celebrity fans scenario to be influenced by different aspects and, consequently, the link to the investigated features being weaker.

We download the Twitter's profile of 6,568,769 unique users that follow at least one of the 21 players for which we were able to find an official Twitter account. From these, only 2,434,183 had a non-empty {\it location} field. We followed the same methodology for the geolocation of these user. We found 540,828 users that match a valid location in Chile. Out of the valid users, only 2,041 had a GPS set of coordinates, and 381,166 did explicit reference to a commune. This gave us a total of 383,207 unique followers that we were able to assign with high confidence to one of the 346 communes in Chile. This is comparable with the 602,810 users that we will use as our sample of followers of the news outlets.

In this new dataset, the number of follower per commune is also strongly correlated with the actual population distribution of Chile ($r(344)=.66, p<.01$). So, it is comparable in size to our newspapers-followers dataset (see Table \ref{table:followers}) and it is a representative sample of the actual Chilean population.

\section{Results}
Our primary task in this work is to approximate the distribution of the audience of the online media based on the geopolitical and socioeconomic characteristics of an area. As discussed above, multiple aspects of a given population can become factors in the news outlets’ audience-targeting strategy. To better understand the media coverage, we first study how this is correlated to geographic elements as predicted by the Gravity Model. To further improve our predictive and explanatory capability, in a second step, we fit a regression model that, besides the geographic feature, also takes into account the political leaning and income level of the communes in the most populated region of the country.





\subsection{Gravity Model}
With the gravity model we want to characterize the news attention as a function of population size and distance from the media source. As we mention before, we model the source as the commune where the headquarter of the news outlet is located or based on the name of the outlet.

We represent each outlet as a vector $F^j$ with the expected number of followers in each commune obtained from the Gravity Model. We also created a vector $T^j$ for each outlet with the actual number of followers on each commune $i$ obtained for our ground-truth. Using the two vectors that represents each outlet we calculate the Pearson correlation coefficients. In Figure~\ref{fig:gm_tw_corr_all} we can see the distribution of the correlation coefficients. We can see that the coverage bias of a big number of outlets can be almost entirely explained just by the geography. Actually, half the outlets correlate over $0.7$. However, there is an important number of outlets for which the geographic bias explains very little or none of their observed coverage.


Table~\ref{table:stats} shows some stats that help to better describe the characteristics of the news outlets with the lowest and highest correlation. Not surprisingly, the group that falls farther from the predicted coverage is dominated by the newspapers in the capital city (i.e., Santiago) and with a national scope. These are expected to be the ones with the most prominent political and socioeconomic bias, given that they are the most influential and the ones that dominate news production. Their leading position also ensures that they receive the biggest share in the investment of advertisers. Hence, these outlets are the most exposed to external pressures. On the other hand, news outlets with a local scope behave as described by the Gravity Model, at least in average. Figures~\ref{fig:gm_tw_corr_local} and~\ref{fig:gm_tw_corr_national} show the distribution of the correlation for the outlets with a local and national scope respectively. The figures illustrate the behavioral difference of these two classes of outlets.

From the previous results, we can conclude that geographic bias is not enough to describe the nature of the news media. If we look for example at the communes \textit{Lo Prado}~\cite{wiki.lo_prado} and \textit{San Miguel}~\cite{wiki.san_miguel}, they have a similar population and are situated at a similar distance from the center of Santiago, where an important number of news outlets (local and national) are located. If we take only these news outlets located in the center of Santiago, the average difference in the expected number of followers between the two communes according to the Gravity Model is just over 1\%. In other words, based only on geographic factors these communes should be virtually indistinguishable. But, if we look at the actual number of followers for the same set of outlets, the average difference is almost 250\%, with an overwhelming dominance of followers from \textit{San Miguel}. Moreover, a general query in the Twitter API for tweets geo-located near ``Lo Prado, Chile'' during August 2017, gives almost 100,000 unique users, while the same query for tweets near ``San Miguel, Chile'' throws only 61,165 unique users. Thus, the difference in news outlets' followers cannot be thought as the result of a disparity in Twitter penetration. One possible factor that may influence this striking contrast is the gap in socioeconomic conditions and deprivation levels between the two communes. \textit{Lo Prado}, despite being located within the capital city, is ranked in the top ten of the poorest communes of Chile~\cite{orellana2017ICVU}. On the other hand, \textit{San Miguel}, even though it is a predominantly residential commune, it is also an important economic/industrial pole of the city. In fact, \textit{San Miguel} ranks in the $40^{th}$ position out of 93 communes in the Index of Urban Quality of Life for Chile~\cite{orellana2017ICVU}. Thus, the hypothesis of our theoretical models in the political economy of the mass medial~\cite{herman1988manufacturing,prat2011political} supports the idea that the socioeconomic characteristics of a sector can make its population more or less attractive to the media.

\subsection{Filtering the data}
Given that we are analyzing geographic coverage and its relation to socioeconomic and political factors, we have to take into account the specific characteristics of Chile. From every point of view, Chile is a heavily centralized country. The previous results detailed in Table \ref{table:stats} give evidence of this. According to a study from 2013~\cite{vonBaer2013descentralizado}, in proportion to its size, population, and economic development, Chile is the most centralized country in Latin America. The data obtained from the INE~\cite{ine.demo} gives us a total estimated population of 17.9 million people for the entire country, out of which 7.4 million (41\%) are located in the Metropolitan Region (where its capital, Santiago, is located). If we also add that this is the smallest (in area) of the 15 regions that compose Chile, we have a very dense population area. Only for its geographic and demographic characteristics, Santiago is already a desirable market for the media based on \textit{Proposition 4} of Prat and Str\"omberg~\cite{prat2011political}. Now, on the political side, each region in Chile is headed by an \textit{Intendente} (equiv. Mayor), but they are appointed and respond directly to the president. Moreover, members of the House of Representatives who legislate on behalf of the different districts of the country reside in Santiago. This organization concentrates almost all the political power in this one region. In the same way, according to the annual report published by the Central Bank of Chile for 2016~\cite{pib.regiones}, the Metropolitan Region participated with 46\% of the GDP (5x the next highest contribution). With this heavily concentrated power in all spheres, and based on our set of hypothesis, the capital of Chile matches all the conditions needed by a population to receive an extensive media coverage.

Consequently, based on our results of the Gravity Model, we will focus on the community of outlets identified as the least influenced by the geographic bias. That is, we filtered our database to keep only those news outlets (locals and national) with headquarter in the capital. The centralization of the Chilean population it is also perceived in our collection of followers: out of 15 regions, 36.9\% of our geolocated followers are in the Metropolitan Region. To minimize the noise in our model, we decided to limit the study of the coverage only to the communes in Santiago. With 51 communes and a wide range of socioeconomic conditions, the Metropolitan Region offers a good case of study on its own. To further strengthen the signal, we also limited the analysis to the 25 news outlets with the highest number of followers.

\subsection{Regression Model}
To extend our model and study the influence of other factors such as the political and socioeconomic characteristics in the distribution of news media followers, we use a regression model. As mention before, our target variable is, given a news outlet and a commune, the ranking position based on the number of followers of that communes for that outlet.

We include three features in our model: {\tt right-leaning}, representing the political dimension; {\tt income}, representing the socioeconomic dimension; and {\tt distance}, representing the geographic dimension. In Table \ref{table:corr} we show the Pearson correlation between our three features (using the filtered data). One thing to notice is the relatively high correlation between the expected income of an average household and the political leaning of the area where it is located. In Chile (and Latin America in general), right-conservative political parties are popularly associated with wealthy people. At least in the last few year, left-leaning parties tend to be more populists.

Using these features, our trained model is able to represent the mass media behavior with high precision. The results of the regression indicate the three predictors explained on average up to 96.3\% $(SD = 0.005)$ of the variance in cross-validation. Figure \ref{fig:learning_curve} shows the learning curve for the selected model.

We were also interested in modeling the coverage behavior of each individual outlet to see how they fit with respect to these three dimensions. To do this, we used the selected features to create a regression model for each news outlet. This model is then used in the same way. That is, we predict their audience-based ranking of the communes in Santiago but using data related only to the selected outlet. The results, shown in Figure \ref{fig:exp_variance_25}, confirm that with the selected features, the regressors are able to approximate the distribution of followers $(M = 0.82, SD = 0.03)$.

We also studied the ranking of the communes in relation to each feature. We used Kendall's Tau (KT) correlations to have an indication of how strong is the influence of each factor in the prediction. Figures \ref{fig:kt_corr_political}, \ref{fig:kt_corr_economic} and \ref{fig:kt_corr_geo} show the distribution of the KT correlation for the top 25 news outlets in Santiago with respect to the communes' political leaning, expected income and distance to the origin, respectively. Results are shown regarding their absolute values because the direction of impact is not important for our model. For example, if a news outlet favors a commune based on the area being right-leaning, for our model this is as telling as another news outlet disregarding the commune for the same reason. In both cases, the outlets are biased based on political factors. The results show that the behavior of news outlets is very similar in terms of the discriminating influence of these three dimensions in the news coverage, at least within this group.

In figure \ref{fig:kt_all_followers} we show a comparison of the KT correlation coefficient for all three dimensions for each of the top 25 news outlets. This comparison can be used as a characterization/profiling of each outlet's coverage behavior. For example, the coverage of Radio Cooperativa (\textit{cooperativa})~\cite{wikiCooperativa} seems to be driven by political and economic factors, with practically no attention to the location of the commune. This is a radio station with a national scope. According to a survey conducted in 2015, it is the second in audience in the region of Santiago~\cite{ipsos.ranking} and the first one among people with the highest income (last quantile). Moreover, its editorial line is ``popularly perceived'' to be associated to the Christian Democratic Party~\cite{wikiCooperativa,Elejalde2018}. Actually, from the early 70's until the late 90's the radio was directly owned by this party (currently belongs to El Mercurio Group). This profile coincides with the characterization reflected by our model. On the other hand, El Quinto Poder (\textit{elquintopoder})~\cite{elquintopoder} is an online news website/community where any member can contribute with its column. This newspaper follows the concept of citizen journalism popularized by sites like \url{http://www.ohmynews.com/}. Its editorial line and community rules explicitly prohibit any content that is aimed at a personal or institutional gain. In the same way, political opinions can only be expressed through personal profiles (rather than an organization profile). In our model, for this newspaper the influence from the political and economic factors are equated, but also the geographic dimensions is the highest within these top 25 outlets.

Just as a comparison, we repeat the analysis, this time filtering the dataset to keep only the top 25 ``newspapers'' in Santiago - i.e., excluding radio station, TV channels, etc. (see figure \ref{fig:kt_all_followers_nps}). Here, for example, it is easy to distinguish newspapers with a local scope, such as \textit{portaldemeli} or \textit{betazeta}. For those, the influence of the geographic factor is higher than that of the economic and even the political features. Actually, in the case of \textit{portaldemeli} (a small commune's local digital newspaper), the economic factor is almost non-existing.

To confirm that the results adhere to the news media ecosystem and do not mimic a behavior common to the social media sphere as a whole, we repeat the experiments on a different topical domain, namely the followers of a group of football players (see section \ref{sec:pol-eco_methodology}). We filtered the data to keep only football players that were born, play or live in Santiago (this condition matched six players). We also kept only the followers that were geolocated in one of the 51 communes of the capital region. The regression model trained with the three selected features, on average, is able to explain only 84\% $(SD = 0.06)$ of the variance in cross validation. Although the model also gives a good fit for this data, it is clearly less explanatory than for the news outlets (over 10\% lost of precision compared to the news outlets). Notice that it is very difficult, if not impossible, to find a public/popular figure for which the followers are not influenced by neither of these three factor. So, the results must be evaluated relative to each other.

Another way to differentiate the two datasets is by comparing the individual influence of each dimension. We calculated the KT correlation coefficient for all three dimensions for each of the top 6 players (see figure \ref{fig:kt_players_vs_outles}). We found that, compared with the news outlets, the difference in the average correlation is statically significant for all three features (right-leaning: $t=8.31, p<.001$; income: $t=7.93, p<.001$; distance $t=2.39, p=.03$).

We can also see that the profile obtained for each individual player (shown in the shaded area in figure \ref{fig:kt_all_followers}) differs from that of the news outlets. In this case, football players tend to have a comparatively stronger influence from the geographic factor and politics plays a lesser role.

The found differences between the two datasets indicate that the distribution of followers for the news outlets is not entirely determined by the social media substrate but is defined by the characteristics of the entities.

In general, our results support the idea of a media system entirely motivated by economic interest as described by the theoretical models that prompted the current study. This profit-driven media system seems to promote selective coverage that targets specific segments of the population based on the ``quality'' of the readership. Notice that a relationship between these features (geographic, social and economic) and media reach is, of course, not a direct proof of a causal relation. However, we assume that there is a natural order of information demand and supply: news media models usually presume that readers get some value from the news they read (e.g., entertainment or arguments to decide on a private action)~\cite{RePEc:cpr:ceprdp:7768,prat2011political,10.1257/aer.96.3.720}. In other words, we consider that people following a newspaper account are interested in the ``editorial line" of that newspaper. This means that the news outlet is creating content that is attractive to a specific audience. 

\section{Conclusions}

This work presents a method to characterize the news outlets in the media system based on the geographic, socioeconomic and political profile of their audiences. Under the assumption of a natural order of information demand and supply (i.e., readers gets some value from the news~\cite{10.1257/aer.96.3.720}), this modeling of the media can imply a conscious targeting of some specific public by catering to their preferences.

Using data from multiple sources we found that news outlets systematically prefer followers from densely populated areas with a specific socioeconomic profile. The political leaning of the commune proved to be the most discriminating feature on the prediction of the level of readership ratings. These findings  support the theoretical claim that describe the news media outlets as profit-driven companies~\cite{herman1988manufacturing,prat2011political}.

Although our model seems to generalize quite well and lends evidence to the hypotheses, we recognize that the methods have some limitations. First, we are restricted to users that we are able to geo-locate using the \textit{location} field from the Twitter profile. A more sophisticated method of location could increase the number of valid users and maybe increase the precision of the model for other regions. A second limitation comes from the fact that the political and economic dimension seem to be closely related. This prevents us from creating a characterization of the outlets that better reflect the actual preference for a population with either a certain political profile or a socioeconomic range, but not both. The entanglement of these two dimensions may be due to the reality of the studied country.

Besides the limitations in the location of the users, the choice to use Twitter followers as a proxy for the audience of the news outlet may introduce some bias and noise to our study. For example, it is difficult to determine the actual demographics of the population in the social media~\cite{Cesare2017DetectionOU}. An alternative method to effectively define the actual audience of the news outlets (e.g., monitoring the passive and active traffic on the selected Twitter accounts or websites) could complement our method and improve the predictive capability of our model in areas with a weaker signal (e.g., beyond the Metropolitan Region). This is left for future work.

In summary, the results seem to support the hypothesis that outlets focus on reaching and acquiring an audience with a higher ``quality'', that can be latter sell to advertisers. This type of media system neglect areas of low population (e.g., rural communes) and high deprivation levels, causing these to be underserved and underrepresented in the news coverage. In turn, this creates a full cycle when public policies and politicians overlook sectors of the population that are less informed and hence, are less likely to influence the \textit{status quo} of the political elite.

We often say that outlets are ``biased''. Thus, it is important to define what it means for these companies to be biased to: one example, and the assumption of our work, is that news outlets are businesses, and as such are motivated by economic interests. If we think of news outlets as a profit-oriented business, we notice that they are perfectly rational in their effort to select content that generates a greater return (a generalization of the ``sex sells'' maxim). It is only when we try to hold on to the traditional view of news outlets as servers of the common good and advocates of democracy that the concept of bias becomes relevant again. Whatever the case, given the influence that news outlets have in society, we should be well aware of their behavior to be able to appropriately process the information they distribute.


\begin{backmatter}
\section{Declarations}
\section*{Availability of data and material}
Twitter's term of service do not allow redistribution of Twitter content but query terms used are included in the manuscript. Using this information, interested researchers can recreate the underlying dataset.

\section*{Competing interests}
  The authors declare that they have no competing interests.
  
\section*{Funding}
The authors acknowledge financial support from Movistar - Telefónica Chile, the Chilean government initiative CORFO 13CEE2-21592 (2013-21592-1-INNOVA\_ PRODUCCION2013-21592-1), Conicyt PAI Networks (REDES170151) ``Geo - Temporal factors in disease spreading and prevention in Chile'' (LF). This work was also supported in part by the doctoral scholarships of CONICYT-PCHA No. 63130228 (EE).

\section*{Author's contributions}
All authors contributed equally to this work.

\section*{Acknowledgements}




\begin{thebibliography}{58}
\ifx \bisbn   \undefined \def \bisbn  #1{ISBN #1}\fi
\ifx \binits  \undefined \def \binits#1{#1}\fi
\ifx \bauthor  \undefined \def \bauthor#1{#1}\fi
\ifx \batitle  \undefined \def \batitle#1{#1}\fi
\ifx \bjtitle  \undefined \def \bjtitle#1{#1}\fi
\ifx \bvolume  \undefined \def \bvolume#1{\textbf{#1}}\fi
\ifx \byear  \undefined \def \byear#1{#1}\fi
\ifx \bissue  \undefined \def \bissue#1{#1}\fi
\ifx \bfpage  \undefined \def \bfpage#1{#1}\fi
\ifx \blpage  \undefined \def \blpage #1{#1}\fi
\ifx \burl  \undefined \def \burl#1{\textsf{#1}}\fi
\ifx \doiurl  \undefined \def \doiurl#1{\textsf{#1}}\fi
\ifx \betal  \undefined \def \betal{\textit{et al.}}\fi
\ifx \binstitute  \undefined \def \binstitute#1{#1}\fi
\ifx \binstitutionaled  \undefined \def \binstitutionaled#1{#1}\fi
\ifx \bctitle  \undefined \def \bctitle#1{#1}\fi
\ifx \beditor  \undefined \def \beditor#1{#1}\fi
\ifx \bpublisher  \undefined \def \bpublisher#1{#1}\fi
\ifx \bbtitle  \undefined \def \bbtitle#1{#1}\fi
\ifx \bedition  \undefined \def \bedition#1{#1}\fi
\ifx \bseriesno  \undefined \def \bseriesno#1{#1}\fi
\ifx \blocation  \undefined \def \blocation#1{#1}\fi
\ifx \bsertitle  \undefined \def \bsertitle#1{#1}\fi
\ifx \bsnm \undefined \def \bsnm#1{#1}\fi
\ifx \bsuffix \undefined \def \bsuffix#1{#1}\fi
\ifx \bparticle \undefined \def \bparticle#1{#1}\fi
\ifx \barticle \undefined \def \barticle#1{#1}\fi
\ifx \bconfdate \undefined \def \bconfdate #1{#1}\fi
\ifx \botherref \undefined \def \botherref #1{#1}\fi
\ifx \url \undefined \def \url#1{\textsf{#1}}\fi
\ifx \bchapter \undefined \def \bchapter#1{#1}\fi
\ifx \bbook \undefined \def \bbook#1{#1}\fi
\ifx \bcomment \undefined \def \bcomment#1{#1}\fi
\ifx \oauthor \undefined \def \oauthor#1{#1}\fi
\ifx \citeauthoryear \undefined \def \citeauthoryear#1{#1}\fi
\ifx \endbibitem  \undefined \def \endbibitem {}\fi
\ifx \bconflocation  \undefined \def \bconflocation#1{#1}\fi
\ifx \arxivurl  \undefined \def \arxivurl#1{\textsf{#1}}\fi
\csname PreBibitemsHook\endcsname

\bibitem{herman1988manufacturing}
\begin{bbook}
\bauthor{\bsnm{Herman}, \binits{E.S.}},
\bauthor{\bsnm{Chomsky}, \binits{N.}}:
\bbtitle{Manufacturing Consent: the Political Economy of the Mass Media},
\bedition{2nd.} edn.
\bpublisher{Pantheon Books},
\blocation{New York, NY}
(\byear{2002}).
\burl{https://books.google.cl/books?id=Up5sAAAAIAAJ}
\end{bbook}
\endbibitem

\bibitem{prat2011political}
\begin{bchapter}
\bauthor{\bsnm{Prat}, \binits{A.}},
\bauthor{\bsnm{Str\"{o}mberg}, \binits{D.}}:
\bctitle{The political economy of mass media}.
In: \beditor{\bsnm{Acemoglu}, \binits{D.}},
\beditor{\bsnm{Arellano}, \binits{M.}},
\beditor{\bsnm{Dekel}, \binits{E.}} (eds.)
\bbtitle{Advances in Economics and Econometrics},
pp. \bfpage{135}--\blpage{187}.
\bpublisher{Cambridge University Press}, \blocation{???}
(\byear{2011}).
\burl{https://doi.org/10.1017/cbo9781139060028.004}
\end{bchapter}
\endbibitem

\bibitem{10.2307/1417611}
\begin{barticle}
\bauthor{\bsnm{Zipf}, \binits{G.K.}}:
\batitle{Some determinants of the circulation of information}.
\bjtitle{The American Journal of Psychology}
\bvolume{59}(\bissue{3}),
\bfpage{401}--\blpage{421}
(\byear{1946})
\end{barticle}
\endbibitem

\bibitem{10.2307/2087063}
\begin{barticle}
\bauthor{\bsnm{Zipf}, \binits{G.K.}}:
\batitle{The {$P_1 P_2/D$} hypothesis: on the intercity movement of persons}.
\bjtitle{American Sociological Review}
\bvolume{11}(\bissue{6}),
\bfpage{677}--\blpage{686}
(\byear{1946})
\end{barticle}
\endbibitem

\bibitem{news.coverage_news.interest_a}
\begin{botherref}
\oauthor{\bsnm{{Pew Research Center for the People \& the Press}}}:
Too much celebrity news, too little good news.
U.S. Politics \& Policy. Pew Research Center
(2007).
Available from:
  {\url{http://www.people-press.org/2007/10/12/too-much-celebrity-news-too-little-good-news/}}
  [Accessed 02-Jan-2018]
\end{botherref}
\endbibitem

\bibitem{news.coverage_news.interest_b}
\begin{botherref}
\oauthor{\bsnm{{Pew Research Center for the People \& the Press}}}:
Haiti, snowstorms, economy vie for public's attention.
The State of the News Media. Journalism \& Media. Pew Research Center
(2010).
Available from:
  {\url{http://www.people-press.org/2010/02/17/haiti-snowstorms-economy-vie-for-publics-attention/}}
  [Accessed 02-Jan-2018]
\end{botherref}
\endbibitem

\bibitem{10.2307/j.ctt7smgs}
\begin{bbook}
\bauthor{\bsnm{Hamilton}, \binits{J.T.}}:
\bbtitle{All the News That's Fit to Sell: How the Market Transforms Information
  Into News}.
\bpublisher{Princeton University Press},
\blocation{Princeton, New Jersey}
(\byear{2004}).
\burl{http://www.jstor.org/stable/j.ctt7smgs}
\end{bbook}
\endbibitem

\bibitem{wiki.native.ads}
\begin{botherref}
\oauthor{\bsnm{Wikipedia}}:
Native Advertising. {Wikipedia, The Free Encyclopedia}.
Available from: {\url{https://en.wikipedia.org/wiki/Native_advertising}
  [Accessed 12-August-2017]}
(2017)
\end{botherref}
\endbibitem

\bibitem{forbes:native-ads}
\begin{botherref}
\oauthor{\bsnm{Fletcher}, \binits{P.}}:
Native advertising will provide a quarter of news media revenue by 2018.
Forbes
(2017).
Available from:
  {\url{https://www.forbes.com/sites/paulfletcher/2016/11/30/native-advertising-will-provide-a-quarter-of-news-media-revenue-by-2018/\#75950afa2d0c}
  [Accessed 12-August-2017]}
\end{botherref}
\endbibitem

\bibitem{doi:10.1162/jeea.2005.3.2-3.259}
\begin{barticle}
\bauthor{\bsnm{Reinikka}, \binits{R.}},
\bauthor{\bsnm{Svensson}, \binits{J.}}:
\batitle{Fighting corruption to improve schooling: Evidence from a newspaper
  campaign in uganda}.
\bjtitle{Journal of the European Economic Association}
\bvolume{3}(\bissue{2-3}),
\bfpage{259}--\blpage{267}
(\byear{2005})
\end{barticle}
\endbibitem

\bibitem{doi:10.1093/wbro/lki002}
\begin{barticle}
\bauthor{\bsnm{Keefer}, \binits{P.}},
\bauthor{\bsnm{Khemani}, \binits{S.}}:
\batitle{Democracy, public expenditures, and the poor: Understanding political
  incentives for providing public services}.
\bjtitle{The World Bank Research Observer}
\bvolume{20}(\bissue{1}),
\bfpage{1}--\blpage{27}
(\byear{2005})
\end{barticle}
\endbibitem

\bibitem{Eagle1029}
\begin{barticle}
\bauthor{\bsnm{Eagle}, \binits{N.}},
\bauthor{\bsnm{Macy}, \binits{M.}},
\bauthor{\bsnm{Claxton}, \binits{R.}}:
\batitle{Network diversity and economic development}.
\bjtitle{Science}
\bvolume{328}(\bissue{5981}),
\bfpage{1029}--\blpage{1031}
(\byear{2010})
\end{barticle}
\endbibitem

\bibitem{Smith:2013:FPI:2441776.2441852}
\begin{bchapter}
\bauthor{\bsnm{Smith}, \binits{C.}},
\bauthor{\bsnm{Quercia}, \binits{D.}},
\bauthor{\bsnm{Capra}, \binits{L.}}:
\bctitle{Finger on the pulse: Identifying deprivation using transit flow
  analysis}.
In: \bbtitle{Proceedings of the 2013 Conference on Computer Supported
  Cooperative Work}.
\bsertitle{CSCW '13},
pp. \bfpage{683}--\blpage{692}.
\bpublisher{ACM},
\blocation{New York, NY, USA}
(\byear{2013}).
\burl{http://doi.acm.org/10.1145/2441776.2441852}
\end{bchapter}
\endbibitem

\bibitem{Park:2012:CFM:2209310.2209311}
\begin{barticle}
\bauthor{\bsnm{Park}, \binits{S.}},
\bauthor{\bsnm{Kang}, \binits{S.}},
\bauthor{\bsnm{Chung}, \binits{S.}},
\bauthor{\bsnm{Song}, \binits{J.}}:
\batitle{A computational framework for media bias mitigation}.
\bjtitle{ACM Trans. Interact. Intell. Syst.}
\bvolume{2}(\bissue{2}),
\bfpage{8}--\blpage{1832}
(\byear{2012})
\end{barticle}
\endbibitem

\bibitem{ECTA:ECTA994}
\begin{barticle}
\bauthor{\bsnm{Gentzkow}, \binits{M.}},
\bauthor{\bsnm{Shapiro}, \binits{J.M.}}:
\batitle{What drives media slant? evidence from u.s. daily newspapers}.
\bjtitle{Econometrica}
\bvolume{78}(\bissue{1}),
\bfpage{35}--\blpage{71}
(\byear{2010})
\end{barticle}
\endbibitem

\bibitem{ICWSM112782}
\begin{bchapter}
\bauthor{\bsnm{Zhou}, \binits{D.X.}},
\bauthor{\bsnm{Resnick}, \binits{P.}},
\bauthor{\bsnm{Mei}, \binits{Q.}}:
\bctitle{Classifying the political leaning of news articles and users from user
  votes}.
In: \bbtitle{Proceedings of the International AAAI Conference on Web and Social
  Media}
(\byear{2011}).
\burl{http://www.aaai.org/ocs/index.php/ICWSM/ICWSM11/paper/view/2782}
\end{bchapter}
\endbibitem

\bibitem{Sudhahar:2012:EDP:2380921.2380938}
\begin{bchapter}
\bauthor{\bsnm{Sudhahar}, \binits{S.}},
\bauthor{\bsnm{Lansdall-Welfare}, \binits{T.}},
\bauthor{\bsnm{Flaounas}, \binits{I.}},
\bauthor{\bsnm{Cristianini}, \binits{N.}}:
\bctitle{Electionwatch: Detecting patterns in news coverage of us elections}.
In: \bbtitle{Proceedings of the Demonstrations at the 13th Conference of the
  European Chapter of the Association for Computational Linguistics}.
\bsertitle{EACL '12},
pp. \bfpage{82}--\blpage{86}.
\bpublisher{Association for Computational Linguistics},
\blocation{Stroudsburg, PA, USA}
(\byear{2012}).
\burl{http://dl.acm.org/citation.cfm?id=2380921.2380938}
\end{bchapter}
\endbibitem

\bibitem{Elejalde2018}
\begin{barticle}
\bauthor{\bsnm{Elejalde}, \binits{E.}},
\bauthor{\bsnm{Ferres}, \binits{L.}},
\bauthor{\bsnm{Herder}, \binits{E.}}:
\batitle{{On the nature of real and perceived bias in the mainstream media}}.
\bjtitle{PLOS ONE}
\bvolume{13}(\bissue{3}),
\bfpage{0193765}
(\byear{2018})
\end{barticle}
\endbibitem

\bibitem{2017arXiv171006347B}
\begin{botherref}
\oauthor{\bsnm{{Bahamonde}}, \binits{J.}},
\oauthor{\bsnm{{Bollen}}, \binits{J.}},
\oauthor{\bsnm{{Elejalde}}, \binits{E.}},
\oauthor{\bsnm{{Ferres}}, \binits{L.}},
\oauthor{\bsnm{{Poblete}}, \binits{B.}}:
{Power Structure in Chilean News Media}.
ArXiv e-prints
(2017).
\arxivurl{1710.06347}
\end{botherref}
\endbibitem

\bibitem{Mullen2009}
\begin{barticle}
\bauthor{\bsnm{Mullen}, \binits{A.}}:
\batitle{The propaganda model after 20 years: Interview with edward s. herman
  and noam chomsky}.
\bjtitle{Westminster Papers in Communication and Culture.}
\bvolume{6}(\bissue{2}),
\bfpage{12}--\blpage{22}
(\byear{2009})
\end{barticle}
\endbibitem

\bibitem{Robinson2015}
\begin{bbook}
\bauthor{\bsnm{Robinson}, \binits{P.}}:
\bbtitle{The Propaganda Model: Still Relevant Today?},
pp. \bfpage{77}--\blpage{96}.
\bpublisher{Palgrave Macmillan UK},
\blocation{London}
(\byear{2015}).
\burl{http://dx.doi.org/10.1007/978-1-137-32021-6\_5}
\end{bbook}
\endbibitem

\bibitem{StateMedia2012}
\begin{botherref}
\oauthor{\bsnm{Edmonds}, \binits{R.}},
\oauthor{\bsnm{Guskin}, \binits{E.}},
\oauthor{\bsnm{Rosenstiel}, \binits{T.}},
\oauthor{\bsnm{Mitchell}, \binits{A.}}:
Newspapers: Building digital revenues proves painfully slow.
The State of the News Media. Journalism \& Media. Pew Research Center
(2012).
Available from:
  {\url{http://assets.pewresearch.org/wp-content/uploads/sites/13/2017/05/24141622/State-of-the-News-Media-Report-2012-FINAL.pdf}
  [Accessed 20-May-2016]}
\end{botherref}
\endbibitem

\bibitem{StateMedia2017a}
\begin{botherref}
\oauthor{\bsnm{Stocking}, \binits{G.}}:
Digital news fact sheet.
The State of the News Media. Journalism \& Media. Pew Research Center
(2017).
Available from: {\url{http://www.journalism.org/fact-sheet/digital-news/}
  [Accessed 20-January-2018]}
\end{botherref}
\endbibitem

\bibitem{StateMedia2017b}
\begin{botherref}
\oauthor{\bsnm{Mitchell}, \binits{A.}},
\oauthor{\bsnm{Gottfried}, \binits{J.}},
\oauthor{\bsnm{Shearer}, \binits{E.}},
\oauthor{\bsnm{Lu}, \binits{K.}}:
How americans encounter, recall and act upon digital news.
Analysis. Journalism \& Media. Pew Research Center
(2017).
Available from:
  {\url{http://www.journalism.org/2017/02/09/how-americans-encounter-recall-and-act-upon-digital-news/}
  [Accessed 20-January-2018]}
\end{botherref}
\endbibitem

\bibitem{StateMedia2017d}
\begin{botherref}
\oauthor{\bsnm{Mitchell}, \binits{A.}},
\oauthor{\bsnm{Gottfried}, \binits{J.}},
\oauthor{\bsnm{Barthel}, \binits{M.}},
\oauthor{\bsnm{Shearer}, \binits{E.}}:
Pathways to news.
The Modern News Consumer. Journalism \& Media. Pew Research Center
(2016).
Available from: {\url{http://www.journalism.org/2016/07/07/pathways-to-news/}
  [Accessed 20-January-2018]}
\end{botherref}
\endbibitem

\bibitem{doi:10.1080/21670811.2015.1133249}
\begin{barticle}
\bauthor{\bsnm{Malik}, \binits{M.M.}},
\bauthor{\bsnm{Pfeffer}, \binits{J.}}:
\batitle{A macroscopic analysis of news content in twitter}.
\bjtitle{Digital Journalism}
\bvolume{4}(\bissue{8}),
\bfpage{955}--\blpage{979}
(\byear{2016})
\end{barticle}
\endbibitem

\bibitem{poderopedia}
\begin{botherref}
\oauthor{\bsnm{{Poderopedia. Poderomedia Foundation}}}:
Mapa de Medios.
Available from: {\url{http://apps.poderopedia.org/mapademedios/index/}
  [Accessed 12-May-2017]}
(2016)
\end{botherref}
\endbibitem

\bibitem{wiki.chilean.newspapers}
\begin{botherref}
\oauthor{\bsnm{Wikipedia}}:
Medios de comunicaci\'on en Chile. {Wikipedia, The Free Encyclopedia}.
Available from:
  {\url{https://es.wikipedia.org/wiki/Medios\_de\_comunicaci\%C3\%B3n\_en\_Chile}
  [Accessed 12-August-2017]}
(2017)
\end{botherref}
\endbibitem

\bibitem{forbes:twitter}
\begin{botherref}
\oauthor{\bsnm{{Trefis Team}}}:
Twitter's surprising user growth bodes well for 2017.
Forbes
(2017).
Available from:
  {\url{https://www.forbes.com/sites/greatspeculations/2017/04/27/twitters-surprising-user-growth-bodes-well-for-2017/\#151668952e11}
  [Accessed 22-August-2017]}
\end{botherref}
\endbibitem

\bibitem{10.1371/journal.pone.0061981}
\begin{barticle}
\bauthor{\bsnm{Mocanu}, \binits{D.}},
\bauthor{\bsnm{Baronchelli}, \binits{A.}},
\bauthor{\bsnm{Perra}, \binits{N.}},
\bauthor{\bsnm{Gon\c{c}alves}, \binits{B.}},
\bauthor{\bsnm{Zhang}, \binits{Q.}},
\bauthor{\bsnm{Vespignani}, \binits{A.}}:
\batitle{The twitter of babel: Mapping world languages through microblogging
  platforms}.
\bjtitle{PLoS ONE}
\bvolume{8}(\bissue{4}),
\bfpage{61981}
(\byear{2013})
\end{barticle}
\endbibitem

\bibitem{ine.demo}
\begin{botherref}
\oauthor{\bsnm{{INE}}}:
Demogr\'aficas vitales. {Instituto Nacional de Estad\'isticas de Chile (INE)}.
Available from:
  {\url{http://www.ine.cl/canales/chile\_estadistico/familias/demograficas\_vitales.php}
  [Accessed 12-December-2017]}
(2013)
\end{botherref}
\endbibitem

\bibitem{pnud2013}
\begin{botherref}
\oauthor{\bsnm{Bravo}, \binits{D.}},
\oauthor{\bsnm{naga}, \binits{O.L.}},
\oauthor{\bsnm{Mill\'an}, \binits{I.}},
\oauthor{\bsnm{Ruiz}, \binits{M.}},
\oauthor{\bsnm{Zamorano}, \binits{F.}}:
Informe final comisi\'on externa revisora del censo 2012.
Technical report,
PNUD,
Santiago, Chile
(2013).
Available from:
  {\url{http://www.cl.undp.org/content/dam/chile/docs/pobreza/undp_cl_pobreza_informe_censo_2013.pdf?download}}
  [Accessed 18-Feb-2018]
\end{botherref}
\endbibitem

\bibitem{casen2013}
\begin{botherref}
\oauthor{\bsnm{Social}, \binits{O.}}:
Encuesta de Caracterizaci\'on Socioecon\'omica Nacional {(CASEN)} 2013.
  {Minesterio de Desarrollo Social (Gobierno de Chile)}.
Available from:
  {\url{http://observatorio.ministeriodesarrollosocial.gob.cl/casen-multidimensional/casen/casen\_2013.php}
  [Accessed 12-December-2017]}
(2015)
\end{botherref}
\endbibitem

\bibitem{servel}
\begin{botherref}
\oauthor{\bsnm{{SERVEL}}}:
Elecciones Presidenciales 1989 al 2013 por Circunscripci\'on Electoral.
  {Servicio Electoral de Chile}.
Available from:
  {\url{https://www.servel.cl/elecciones-presidenciales-1989-al-2013-por-circunscripcion-electoral/}
  [Accessed 12-August-2017]}
(2017)
\end{botherref}
\endbibitem

\bibitem{Ozdikis:2017:SLE:3125062.3125098}
\begin{barticle}
\bauthor{\bsnm{Ozdikis}, \binits{O.}},
\bauthor{\bsnm{O\u{g}uzt\"{u}z\"{u}n}, \binits{H.}},
\bauthor{\bsnm{Karagoz}, \binits{P.}}:
\batitle{A survey on location estimation techniques for events detected in
  twitter}.
\bjtitle{Knowl. Inf. Syst.}
\bvolume{52}(\bissue{2}),
\bfpage{291}--\blpage{339}
(\byear{2017})
\end{barticle}
\endbibitem

\bibitem{Ajao:2015:SLI:2879214.2879221}
\begin{barticle}
\bauthor{\bsnm{Ajao}, \binits{O.}},
\bauthor{\bsnm{Hong}, \binits{J.}},
\bauthor{\bsnm{Liu}, \binits{W.}}:
\batitle{A survey of location inference techniques on twitter}.
\bjtitle{J. Inf. Sci.}
\bvolume{41}(\bissue{6}),
\bfpage{855}--\blpage{864}
(\byear{2015})
\end{barticle}
\endbibitem

\bibitem{7998610}
\begin{barticle}
\bauthor{\bsnm{Kordopatis-Zilos}, \binits{G.}},
\bauthor{\bsnm{Papadopoulos}, \binits{S.}},
\bauthor{\bsnm{Kompatsiaris}, \binits{I.}}:
\batitle{Geotagging text content with language models and feature mining}.
\bjtitle{Proceedings of the {IEEE}}
\bvolume{105}(\bissue{10}),
\bfpage{1971}--\blpage{1986}
(\byear{2017})
\end{barticle}
\endbibitem

\bibitem{Amitay:2004:WGW:1008992.1009040}
\begin{bchapter}
\bauthor{\bsnm{Amitay}, \binits{E.}},
\bauthor{\bsnm{Har'El}, \binits{N.}},
\bauthor{\bsnm{Sivan}, \binits{R.}},
\bauthor{\bsnm{Soffer}, \binits{A.}}:
\bctitle{Web-a-where: Geotagging web content}.
In: \bbtitle{Proceedings of the 27th Annual International ACM SIGIR Conference
  on Research and Development in Information Retrieval}.
\bsertitle{SIGIR '04},
pp. \bfpage{273}--\blpage{280}.
\bpublisher{ACM},
\blocation{New York, NY, USA}
(\byear{2004}).
\burl{http://doi.acm.org/10.1145/1008992.1009040}
\end{bchapter}
\endbibitem

\bibitem{Serdyukov:2009:PFP:1571941.1572025}
\begin{bchapter}
\bauthor{\bsnm{Serdyukov}, \binits{P.}},
\bauthor{\bsnm{Murdock}, \binits{V.}},
\bauthor{\bparticle{van} \bsnm{Zwol}, \binits{R.}}:
\bctitle{Placing flickr photos on a map}.
In: \bbtitle{Proceedings of the 32Nd International ACM SIGIR Conference on
  Research and Development in Information Retrieval}.
\bsertitle{SIGIR '09},
pp. \bfpage{484}--\blpage{491}.
\bpublisher{ACM},
\blocation{New York, NY, USA}
(\byear{2009}).
\burl{http://doi.acm.org/10.1145/1571941.1572025}
\end{bchapter}
\endbibitem

\bibitem{Cheng:2010:YYT:1871437.1871535}
\begin{bchapter}
\bauthor{\bsnm{Cheng}, \binits{Z.}},
\bauthor{\bsnm{Caverlee}, \binits{J.}},
\bauthor{\bsnm{Lee}, \binits{K.}}:
\bctitle{You are where you tweet: A content-based approach to geo-locating
  twitter users}.
In: \bbtitle{Proceedings of the 19th ACM International Conference on
  Information and Knowledge Management}.
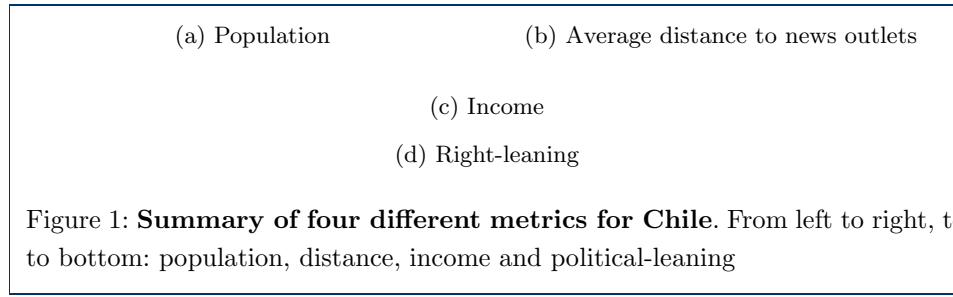
\bsertitle{CIKM '10},
pp. \bfpage{759}--\blpage{768}.
\bpublisher{ACM},
\blocation{New York, NY, USA}
(\byear{2010}).
\burl{http://doi.acm.org/10.1145/1871437.1871535}
\end{bchapter}
\endbibitem

\bibitem{Ryoo:2014:ITU:2567948.2579236}
\begin{bchapter}
\bauthor{\bsnm{Ryoo}, \binits{K.}},
\bauthor{\bsnm{Moon}, \binits{S.}}:
\bctitle{Inferring twitter user locations with 10 km accuracy}.
In: \bbtitle{Proceedings of the 23rd International Conference on World Wide
  Web}.
\bsertitle{WWW '14 Companion},
pp. \bfpage{643}--\blpage{648}.
\bpublisher{ACM},
\blocation{New York, NY, USA}
(\byear{2014}).
\burl{http://doi.acm.org/10.1145/2567948.2579236}
\end{bchapter}
\endbibitem

\bibitem{Backstrom:2010:FMY:1772690.1772698}
\begin{bchapter}
\bauthor{\bsnm{Backstrom}, \binits{L.}},
\bauthor{\bsnm{Sun}, \binits{E.}},
\bauthor{\bsnm{Marlow}, \binits{C.}}:
\bctitle{Find me if you can: Improving geographical prediction with social and
  spatial proximity}.
In: \bbtitle{Proceedings of the 19th International Conference on World Wide
  Web}.
\bsertitle{WWW '10},
pp. \bfpage{61}--\blpage{70}.
\bpublisher{ACM},
\blocation{New York, NY, USA}
(\byear{2010}).
\burl{http://doi.acm.org/10.1145/1772690.1772698}
\end{bchapter}
\endbibitem

\bibitem{Kong:2014:SLS:2733004.2733060}
\begin{barticle}
\bauthor{\bsnm{Kong}, \binits{L.}},
\bauthor{\bsnm{Liu}, \binits{Z.}},
\bauthor{\bsnm{Huang}, \binits{Y.}}:
\batitle{Spot: Locating social media users based on social network context}.
\bjtitle{Proc. VLDB Endow.}
\bvolume{7}(\bissue{13}),
\bfpage{1681}--\blpage{1684}
(\byear{2014})
\end{barticle}
\endbibitem

\bibitem{Hecht:2011:TJB:1978942.1978976}
\begin{bchapter}
\bauthor{\bsnm{Hecht}, \binits{B.}},
\bauthor{\bsnm{Hong}, \binits{L.}},
\bauthor{\bsnm{Suh}, \binits{B.}},
\bauthor{\bsnm{Chi}, \binits{E.H.}}:
\bctitle{Tweets from justin bieber's heart: The dynamics of the location field
  in user profiles}.
In: \bbtitle{Proceedings of the SIGCHI Conference on Human Factors in Computing
  Systems}.
\bsertitle{CHI '11},
pp. \bfpage{237}--\blpage{246}.
\bpublisher{ACM},
\blocation{New York, NY, USA}
(\byear{2011}).
\burl{http://doi.acm.org/10.1145/1978942.1978976}
\end{bchapter}
\endbibitem

\bibitem{rulamahue}
\begin{botherref}
\oauthor{\bsnm{Albers}, \binits{C.}}:
Mapas de las Provincias y Regiones de Chile. {Cartograf\'ia Rulamahue}.
Available from: {\url{http://www.rulamahue.cl/}} [Accessed 18-October-2017]
(2016)
\end{botherref}
\endbibitem

\bibitem{hallin2004comparing}
\begin{bbook}
\bauthor{\bsnm{Hallin}, \binits{D.C.}},
\bauthor{\bsnm{Mancini}, \binits{P.}}:
\bbtitle{Comparing Media Systems: Three Models of Media and Politics}.
\bsertitle{Communication, Society and Politics}.
\bpublisher{Cambridge University Press}, \blocation{???}
(\byear{2004}).
\burl{https://books.google.it/books?id=954NJChZAGoC}
\end{bbook}
\endbibitem

\bibitem{Breiman:2001:RF:570181.570182}
\begin{barticle}
\bauthor{\bsnm{Breiman}, \binits{L.}}:
\batitle{Random forests}.
\bjtitle{Mach. Learn.}
\bvolume{45}(\bissue{1}),
\bfpage{5}--\blpage{32}
(\byear{2001})
\end{barticle}
\endbibitem

\bibitem{wiki.lo_prado}
\begin{botherref}
\oauthor{\bsnm{Wikipedia}}:
Lo Prado. {Wikipedia, The Free Encyclopedia}.
Available from: {\url{https://es.wikipedia.org/wiki/Lo\_Prado} [Accessed
  12-August-2017]}
(2017)
\end{botherref}
\endbibitem

\bibitem{wiki.san_miguel}
\begin{botherref}
\oauthor{\bsnm{Wikipedia}}:
San Miguel (Chile). {Wikipedia, The Free Encyclopedia}.
Available from: {\url{https://es.wikipedia.org/wiki/San\_Miguel\_(Chile)}
  [Accessed 12-August-2017]}
(2017)
\end{botherref}
\endbibitem

\bibitem{orellana2017ICVU}
\begin{botherref}
\oauthor{\bsnm{Orellana}, \binits{A.}}:
Icvu 2017: \'indice de calidad de vida urbana. comunas y ciudades de chile.
Technical report,
Instituto de Estudios Urbanos y Territoriales. Pontificia Universidad
  Cat\'olica de Chile,
Santiago, Chile
(2017).
Available from:
  {\url{http://fadeu.uc.cl/images/noticias/2017/05.Mayo/Presentacion_ICVU_2017_.pdf}}
  [Accessed 10-January-2018]
\end{botherref}
\endbibitem

\bibitem{vonBaer2013descentralizado}
\begin{botherref}
\oauthor{\bparticle{von} \bsnm{Baer}, \binits{H.}},
\oauthor{\bsnm{Torralbo}, \binits{F.}}:
Chile descentralizado y desarrollado: Fundamentos y propuestas para construir
  una pol\'itica de estado en descentralizaci\'on y desarrollo territorial en
  chile.
95 propuestas para un Chile mejor
(2013).
Available from:
  {\url{http://95propuestas.cl/site/wp-content/uploads/2013/05/chile-descentralizado-y-desarrollado-heinrich-von-baer-y-felipe-torralbo.pdf}}
  [Accessed 10-January-2018]
\end{botherref}
\endbibitem

\bibitem{pib.regiones}
\begin{botherref}
\oauthor{\bparticle{de} \bsnm{Chile}, \binits{B.C.}}:
Cuentas Nacionales de Chile. {PIB} Regional 2016.
Available from:
  \url{http://www.bcentral.cl/documents/20143/32019/CCNNPIB_Regional2016.pdf}
  [Accessed 30-December-2017]
(2017)
\end{botherref}
\endbibitem

\bibitem{wikiCooperativa}
\begin{botherref}
\oauthor{\bsnm{Wikipedia}}:
Radio Cooperativa (Chile). {Wikipedia, The Free Encyclopedia}.
Available from:
  {\url{https://es.wikipedia.org/wiki/Radio\_Cooperativa\_(Chile)} [Accessed
  30-December-2017]}
(2017)
\end{botherref}
\endbibitem

\bibitem{ipsos.ranking}
\begin{botherref}
\oauthor{\bsnm{{Ipsos}}}:
Ranking General de Audiencia Gran Santiago. Ipsos Radio 2016.
Available from: {\url{http://www.ipsos.cl/ipsosradioalaire/pagdos.htm}}
  [Accessed 30-December-2017]
(2015)
\end{botherref}
\endbibitem

\bibitem{elquintopoder}
\begin{botherref}
\oauthor{\bsnm{{Fundaci\'on Democracia y Desarrollo}}}:
El Quinto Poder. {FD+D}.
Available from: {\url{http://www.elquintopoder.cl/} [Accessed
  30-December-2017]}
(2010)
\end{botherref}
\endbibitem

\bibitem{RePEc:cpr:ceprdp:7768}
\begin{barticle}
\bauthor{\bsnm{Anderson}, \binits{S.P.}},
\bauthor{\bsnm{McLaren}, \binits{J.}}:
\batitle{Media mergers and media bias with rational consumers}.
\bjtitle{Journal of the European Economic Association}
\bvolume{10}(\bissue{4}),
\bfpage{831}--\blpage{859}
(\byear{2012})
\end{barticle}
\endbibitem

\bibitem{10.1257/aer.96.3.720}
\begin{barticle}
\bauthor{\bsnm{Besley}, \binits{T.}},
\bauthor{\bsnm{Prat}, \binits{A.}}:
\batitle{Handcuffs for the grabbing hand? media capture and government
  accountability}.
\bjtitle{American Economic Review}
\bvolume{96}(\bissue{3}),
\bfpage{720}--\blpage{736}
(\byear{2006})
\end{barticle}
\endbibitem

\bibitem{Cesare2017DetectionOU}
\begin{botherref}
\oauthor{\bsnm{Cesare}, \binits{N.}},
\oauthor{\bsnm{Grant}, \binits{C.}},
\oauthor{\bsnm{Nsoesie}, \binits{E.O.}}:
Detection of user demographics on social media: A review of methods and
  recommendations for best practices.
CoRR
\textbf{abs/1702.01807}
(2017)
\end{botherref}
\endbibitem

\end{thebibliography}

\newcommand{\BMCxmlcomment}[1]{}

\BMCxmlcomment{

<refgrp>

<bibl id="B1">
  <title><p>Manufacturing Consent: the Political Economy of the Mass
  Media</p></title>
  <aug>
    <au><snm>Herman</snm><fnm>ES</fnm></au>
    <au><snm>Chomsky</snm><fnm>N</fnm></au>
  </aug>
  <publisher>New York, NY: Pantheon Books</publisher>
  <edition>2</edition>
  <pubdate>2002</pubdate>
  <url>https://books.google.cl/books?id=Up5sAAAAIAAJ</url>
</bibl>

<bibl id="B2">
  <title><p>The Political Economy of Mass Media</p></title>
  <aug>
    <au><snm>Prat</snm><fnm>A</fnm></au>
    <au><snm>Str\"{o}mberg</snm><fnm>D</fnm></au>
  </aug>
  <source>Advances in Economics and Econometrics</source>
  <publisher>Cambridge University Press</publisher>
  <editor>Daron Acemoglu and Manuel Arellano and Eddie Dekel</editor>
  <pubdate>2011</pubdate>
  <fpage>135</fpage>
  <lpage>-187</lpage>
  <url>https://doi.org/10.1017/cbo9781139060028.004</url>
</bibl>

<bibl id="B3">
  <title><p>Some Determinants of the Circulation of Information</p></title>
  <aug>
    <au><snm>Zipf</snm><fnm>GK</fnm></au>
  </aug>
  <source>The American Journal of Psychology</source>
  <publisher>University of Illinois Press</publisher>
  <pubdate>1946</pubdate>
  <volume>59</volume>
  <issue>3</issue>
  <fpage>401</fpage>
  <lpage>421</lpage>
  <url>http://www.jstor.org/stable/1417611</url>
</bibl>

<bibl id="B4">
  <title><p>The {$P_1 P_2/D$} Hypothesis: on the Intercity Movement of
  Persons</p></title>
  <aug>
    <au><snm>Zipf</snm><fnm>GK</fnm></au>
  </aug>
  <source>American Sociological Review</source>
  <publisher>[American Sociological Association, Sage Publications,
  Inc.]</publisher>
  <pubdate>1946</pubdate>
  <volume>11</volume>
  <issue>6</issue>
  <fpage>677</fpage>
  <lpage>686</lpage>
  <url>http://www.jstor.org/stable/2087063</url>
</bibl>

<bibl id="B5">
  <title><p>Too Much Celebrity News, too Little Good News</p></title>
  <aug>
    <au><cnm>{Pew Research Center for the People \& the Press}</cnm></au>
  </aug>
  <source>U.S. Politics \& Policy. Pew Research Center</source>
  <pubdate>2007</pubdate>
  <note>Available from:
  {\url{http://www.people-press.org/2007/10/12/too-much-celebrity-news-too-little-good-news/}}
  [Accessed 02-Jan-2018]</note>
</bibl>

<bibl id="B6">
  <title><p>Haiti, Snowstorms, Economy Vie for Public's Attention</p></title>
  <aug>
    <au><cnm>{Pew Research Center for the People \& the Press}</cnm></au>
  </aug>
  <source>The State of the News Media. Journalism \& Media. Pew Research
  Center</source>
  <pubdate>2010</pubdate>
  <note>Available from:
  {\url{http://www.people-press.org/2010/02/17/haiti-snowstorms-economy-vie-for-publics-attention/}}
  [Accessed 02-Jan-2018]</note>
</bibl>

<bibl id="B7">
  <title><p>All the News that's Fit to Sell: How the Market Transforms
  Information into News</p></title>
  <aug>
    <au><snm>Hamilton</snm><fnm>JT</fnm></au>
  </aug>
  <publisher>Princeton, New Jersey: Princeton University Press</publisher>
  <pubdate>2004</pubdate>
  <url>http://www.jstor.org/stable/j.ctt7smgs</url>
</bibl>

<bibl id="B8">
  <title><p>Native Advertising. {Wikipedia, The Free Encyclopedia}</p></title>
  <aug>
    <au><cnm>Wikipedia</cnm></au>
  </aug>
  <pubdate>2017</pubdate>
  <note>Available from: {\url{https://en.wikipedia.org/wiki/Native_advertising}
  [Accessed 12-August-2017]}</note>
</bibl>

<bibl id="B9">
  <title><p>Native Advertising will Provide a Quarter of News Media Revenue by
  2018</p></title>
  <aug>
    <au><snm>Fletcher</snm><fnm>P</fnm></au>
  </aug>
  <source>Forbes</source>
  <pubdate>2017</pubdate>
  <note>Available from:
  {\url{https://www.forbes.com/sites/paulfletcher/2016/11/30/native-advertising-will-provide-a-quarter-of-news-media-revenue-by-2018/\#75950afa2d0c}
  [Accessed 12-August-2017]}</note>
</bibl>

<bibl id="B10">
  <title><p>Fighting Corruption to Improve Schooling: Evidence from a Newspaper
  Campaign in Uganda</p></title>
  <aug>
    <au><snm>Reinikka</snm><fnm>R</fnm></au>
    <au><snm>Svensson</snm><fnm>J</fnm></au>
  </aug>
  <source>Journal of the European Economic Association</source>
  <pubdate>2005</pubdate>
  <volume>3</volume>
  <issue>2-3</issue>
  <fpage>259</fpage>
  <lpage>267</lpage>
  <url>http://dx.doi.org/10.1162/jeea.2005.3.2-3.259</url>
</bibl>

<bibl id="B11">
  <title><p>Democracy, Public Expenditures, and the Poor: Understanding
  Political Incentives for Providing Public Services</p></title>
  <aug>
    <au><snm>Keefer</snm><fnm>P</fnm></au>
    <au><snm>Khemani</snm><fnm>S</fnm></au>
  </aug>
  <source>The World Bank Research Observer</source>
  <pubdate>2005</pubdate>
  <volume>20</volume>
  <issue>1</issue>
  <fpage>1</fpage>
  <lpage>27</lpage>
  <url>http://dx.doi.org/10.1093/wbro/lki002</url>
</bibl>

<bibl id="B12">
  <title><p>Network Diversity and Economic Development</p></title>
  <aug>
    <au><snm>Eagle</snm><fnm>N</fnm></au>
    <au><snm>Macy</snm><fnm>M</fnm></au>
    <au><snm>Claxton</snm><fnm>R</fnm></au>
  </aug>
  <source>Science</source>
  <publisher>American Association for the Advancement of Science</publisher>
  <pubdate>2010</pubdate>
  <volume>328</volume>
  <issue>5981</issue>
  <fpage>1029</fpage>
  <lpage>-1031</lpage>
  <url>http://science.sciencemag.org/content/328/5981/1029</url>
</bibl>

<bibl id="B13">
  <title><p>Finger on the Pulse: Identifying Deprivation Using Transit Flow
  Analysis</p></title>
  <aug>
    <au><snm>Smith</snm><fnm>C</fnm></au>
    <au><snm>Quercia</snm><fnm>D</fnm></au>
    <au><snm>Capra</snm><fnm>L</fnm></au>
  </aug>
  <source>Proceedings of the 2013 Conference on Computer Supported Cooperative
  Work</source>
  <publisher>New York, NY, USA: ACM</publisher>
  <series><title><p>CSCW '13</p></title></series>
  <pubdate>2013</pubdate>
  <fpage>683</fpage>
  <lpage>-692</lpage>
  <url>http://doi.acm.org/10.1145/2441776.2441852</url>
</bibl>

<bibl id="B14">
  <title><p>A Computational Framework for Media Bias Mitigation</p></title>
  <aug>
    <au><snm>Park</snm><fnm>S</fnm></au>
    <au><snm>Kang</snm><fnm>S</fnm></au>
    <au><snm>Chung</snm><fnm>S</fnm></au>
    <au><snm>Song</snm><fnm>J</fnm></au>
  </aug>
  <source>ACM Trans. Interact. Intell. Syst.</source>
  <publisher>New York, NY, USA: ACM</publisher>
  <pubdate>2012</pubdate>
  <volume>2</volume>
  <issue>2</issue>
  <fpage>8:1</fpage>
  <lpage>-8:32</lpage>
  <url>http://doi.acm.org/10.1145/2209310.2209311</url>
</bibl>

<bibl id="B15">
  <title><p>What Drives Media Slant? Evidence from U.S. Daily
  Newspapers</p></title>
  <aug>
    <au><snm>Gentzkow</snm><fnm>M</fnm></au>
    <au><snm>Shapiro</snm><fnm>JM</fnm></au>
  </aug>
  <source>Econometrica</source>
  <publisher>Blackwell Publishing Ltd</publisher>
  <pubdate>2010</pubdate>
  <volume>78</volume>
  <issue>1</issue>
  <fpage>35</fpage>
  <lpage>-71</lpage>
  <url>http://dx.doi.org/10.3982/ECTA7195</url>
</bibl>

<bibl id="B16">
  <title><p>Classifying the Political Leaning of News Articles and Users from
  User Votes</p></title>
  <aug>
    <au><snm>Zhou</snm><fnm>DX</fnm></au>
    <au><snm>Resnick</snm><fnm>P</fnm></au>
    <au><snm>Mei</snm><fnm>Q</fnm></au>
  </aug>
  <source>Proceedings of the International AAAI Conference on Web and Social
  Media</source>
  <pubdate>2011</pubdate>
  <url>http://www.aaai.org/ocs/index.php/ICWSM/ICWSM11/paper/view/2782</url>
</bibl>

<bibl id="B17">
  <title><p>ElectionWatch: Detecting Patterns in News Coverage of US
  Elections</p></title>
  <aug>
    <au><snm>Sudhahar</snm><fnm>S</fnm></au>
    <au><snm>Lansdall Welfare</snm><fnm>T</fnm></au>
    <au><snm>Flaounas</snm><fnm>I</fnm></au>
    <au><snm>Cristianini</snm><fnm>N</fnm></au>
  </aug>
  <source>Proceedings of the Demonstrations at the 13th Conference of the
  European Chapter of the Association for Computational Linguistics</source>
  <publisher>Stroudsburg, PA, USA: Association for Computational
  Linguistics</publisher>
  <series><title><p>EACL '12</p></title></series>
  <pubdate>2012</pubdate>
  <fpage>82</fpage>
  <lpage>-86</lpage>
  <url>http://dl.acm.org/citation.cfm?id=2380921.2380938</url>
</bibl>

<bibl id="B18">
  <title><p>{On the nature of real and perceived bias in the mainstream
  media}</p></title>
  <aug>
    <au><snm>Elejalde</snm><fnm>E</fnm></au>
    <au><snm>Ferres</snm><fnm>L</fnm></au>
    <au><snm>Herder</snm><fnm>E</fnm></au>
  </aug>
  <source>PLOS ONE</source>
  <publisher>Public Library of Science</publisher>
  <editor>Chialvo, Dante R.</editor>
  <pubdate>2018</pubdate>
  <volume>13</volume>
  <issue>3</issue>
  <fpage>e0193765</fpage>
</bibl>

<bibl id="B19">
  <title><p>{Power Structure in Chilean News Media}</p></title>
  <aug>
    <au><snm>{Bahamonde}</snm><fnm>J.</fnm></au>
    <au><snm>{Bollen}</snm><fnm>J.</fnm></au>
    <au><snm>{Elejalde}</snm><fnm>E.</fnm></au>
    <au><snm>{Ferres}</snm><fnm>L.</fnm></au>
    <au><snm>{Poblete}</snm><fnm>B.</fnm></au>
  </aug>
  <source>ArXiv e-prints</source>
  <pubdate>2017</pubdate>
</bibl>

<bibl id="B20">
  <title><p>The Propaganda Model after 20 Years: Interview with Edward S.
  Herman and Noam Chomsky</p></title>
  <aug>
    <au><snm>Mullen</snm><fnm>A</fnm></au>
  </aug>
  <source>Westminster Papers in Communication and Culture.</source>
  <pubdate>2009</pubdate>
  <volume>6</volume>
  <issue>2</issue>
  <fpage>12</fpage>
  <lpage>-22</lpage>
  <url>http://doi.org/10.16997/wpcc.121</url>
</bibl>

<bibl id="B21">
  <title><p>The Propaganda Model: Still Relevant Today?</p></title>
  <aug>
    <au><snm>Robinson</snm><fnm>P</fnm></au>
  </aug>
  <source>Noam Chomsky</source>
  <publisher>London: Palgrave Macmillan UK</publisher>
  <pubdate>2015</pubdate>
  <fpage>77</fpage>
  <lpage>-96</lpage>
  <url>http://dx.doi.org/10.1007/978-1-137-32021-6\_5</url>
</bibl>

<bibl id="B22">
  <title><p>Newspapers: Building Digital Revenues Proves Painfully
  Slow</p></title>
  <aug>
    <au><snm>Edmonds</snm><fnm>R</fnm></au>
    <au><snm>Guskin</snm><fnm>E</fnm></au>
    <au><snm>Rosenstiel</snm><fnm>T</fnm></au>
    <au><snm>Mitchell</snm><fnm>A</fnm></au>
  </aug>
  <source>The State of the News Media. Journalism \& Media. Pew Research
  Center</source>
  <pubdate>2012</pubdate>
  <note>Available from:
  {\url{http://assets.pewresearch.org/wp-content/uploads/sites/13/2017/05/24141622/State-of-the-News-Media-Report-2012-FINAL.pdf}
  [Accessed 20-May-2016]}</note>
</bibl>

<bibl id="B23">
  <title><p>Digital News Fact Sheet</p></title>
  <aug>
    <au><snm>Stocking</snm><fnm>G</fnm></au>
  </aug>
  <source>The State of the News Media. Journalism \& Media. Pew Research
  Center</source>
  <pubdate>2017</pubdate>
  <note>Available from:
  {\url{http://www.journalism.org/fact-sheet/digital-news/} [Accessed
  20-January-2018]}</note>
</bibl>

<bibl id="B24">
  <title><p>How Americans Encounter, Recall and Act Upon Digital
  News</p></title>
  <aug>
    <au><snm>Mitchell</snm><fnm>A</fnm></au>
    <au><snm>Gottfried</snm><fnm>J</fnm></au>
    <au><snm>Shearer</snm><fnm>E</fnm></au>
    <au><snm>Lu</snm><fnm>K</fnm></au>
  </aug>
  <source>Analysis. Journalism \& Media. Pew Research Center</source>
  <pubdate>2017</pubdate>
  <note>Available from:
  {\url{http://www.journalism.org/2017/02/09/how-americans-encounter-recall-and-act-upon-digital-news/}
  [Accessed 20-January-2018]}</note>
</bibl>

<bibl id="B25">
  <title><p>Pathways to news</p></title>
  <aug>
    <au><snm>Mitchell</snm><fnm>A</fnm></au>
    <au><snm>Gottfried</snm><fnm>J</fnm></au>
    <au><snm>Barthel</snm><fnm>M</fnm></au>
    <au><snm>Shearer</snm><fnm>E</fnm></au>
  </aug>
  <source>The Modern News Consumer. Journalism \& Media. Pew Research
  Center</source>
  <pubdate>2016</pubdate>
  <note>Available from:
  {\url{http://www.journalism.org/2016/07/07/pathways-to-news/} [Accessed
  20-January-2018]}</note>
</bibl>

<bibl id="B26">
  <title><p>A Macroscopic Analysis of News Content in Twitter</p></title>
  <aug>
    <au><snm>Malik</snm><fnm>MM</fnm></au>
    <au><snm>Pfeffer</snm><fnm>J</fnm></au>
  </aug>
  <source>Digital Journalism</source>
  <pubdate>2016</pubdate>
  <volume>4</volume>
  <issue>8</issue>
  <fpage>955</fpage>
  <lpage>979</lpage>
  <url>http://dx.doi.org/10.1080/21670811.2015.1133249</url>
</bibl>

<bibl id="B27">
  <title><p>Mapa de Medios</p></title>
  <aug>
    <au><cnm>{Poderopedia. Poderomedia Foundation}</cnm></au>
  </aug>
  <pubdate>2016</pubdate>
  <note>Available from: {\url{http://apps.poderopedia.org/mapademedios/index/}
  [Accessed 12-May-2017]}</note>
</bibl>

<bibl id="B28">
  <title><p>Medios de comunicaci\'on en Chile. {Wikipedia, The Free
  Encyclopedia}</p></title>
  <aug>
    <au><cnm>Wikipedia</cnm></au>
  </aug>
  <pubdate>2017</pubdate>
  <note>Available from:
  {\url{https://es.wikipedia.org/wiki/Medios\_de\_comunicaci\%C3\%B3n\_en\_Chile}
  [Accessed 12-August-2017]}</note>
</bibl>

<bibl id="B29">
  <title><p>Twitter's Surprising User Growth Bodes Well For 2017</p></title>
  <aug>
    <au><cnm>{Trefis Team}</cnm></au>
  </aug>
  <source>Forbes</source>
  <pubdate>2017</pubdate>
  <note>Available from:
  {\url{https://www.forbes.com/sites/greatspeculations/2017/04/27/twitters-surprising-user-growth-bodes-well-for-2017/\#151668952e11}
  [Accessed 22-August-2017]}</note>
</bibl>

<bibl id="B30">
  <title><p>The Twitter of Babel: Mapping World Languages through Microblogging
  Platforms</p></title>
  <aug>
    <au><snm>Mocanu</snm><fnm>D</fnm></au>
    <au><snm>Baronchelli</snm><fnm>A</fnm></au>
    <au><snm>Perra</snm><fnm>N</fnm></au>
    <au><snm>Gon\c{c}alves</snm><fnm>B</fnm></au>
    <au><snm>Zhang</snm><fnm>Q</fnm></au>
    <au><snm>Vespignani</snm><fnm>A</fnm></au>
  </aug>
  <source>PLoS ONE</source>
  <publisher>Public Library of Science</publisher>
  <pubdate>2013</pubdate>
  <volume>8</volume>
  <issue>4</issue>
  <fpage>e61981</fpage>
  <url>http://dx.doi.org/10.1371\%2Fjournal.pone.0061981</url>
</bibl>

<bibl id="B31">
  <title><p>Demogr\'aficas vitales. {Instituto Nacional de Estad\'isticas de
  Chile (INE)}</p></title>
  <aug>
    <au><cnm>{INE}</cnm></au>
  </aug>
  <pubdate>2013</pubdate>
  <note>Available from:
  {\url{http://www.ine.cl/canales/chile\_estadistico/familias/demograficas\_vitales.php}
  [Accessed 12-December-2017]}</note>
</bibl>

<bibl id="B32">
  <title><p>Informe final comisi\'on externa revisora del CENSO
  2012</p></title>
  <aug>
    <au><snm>Bravo</snm><fnm>D</fnm></au>
    <au><snm>naga</snm><fnm>OL</fnm></au>
    <au><snm>Mill\'an</snm><fnm>I</fnm></au>
    <au><snm>Ruiz</snm><fnm>M</fnm></au>
    <au><snm>Zamorano</snm><fnm>F</fnm></au>
  </aug>
  <publisher>Santiago, Chile</publisher>
  <pubdate>2013</pubdate>
  <note>Available from:
  {\url{http://www.cl.undp.org/content/dam/chile/docs/pobreza/undp_cl_pobreza_informe_censo_2013.pdf?download}}
  [Accessed 18-Feb-2018]</note>
</bibl>

<bibl id="B33">
  <title><p>Encuesta de Caracterizaci\'on Socioecon\'omica Nacional {(CASEN)}
  2013. {Minesterio de Desarrollo Social (Gobierno de Chile)}</p></title>
  <aug>
    <au><snm>Social</snm><fnm>O</fnm></au>
  </aug>
  <pubdate>2015</pubdate>
  <note>Available from:
  {\url{http://observatorio.ministeriodesarrollosocial.gob.cl/casen-multidimensional/casen/casen\_2013.php}
  [Accessed 12-December-2017]}</note>
</bibl>

<bibl id="B34">
  <title><p>Elecciones Presidenciales 1989 al 2013 por Circunscripci\'on
  Electoral. {Servicio Electoral de Chile}</p></title>
  <aug>
    <au><cnm>{SERVEL}</cnm></au>
  </aug>
  <pubdate>2017</pubdate>
  <note>Available from:
  {\url{https://www.servel.cl/elecciones-presidenciales-1989-al-2013-por-circunscripcion-electoral/}
  [Accessed 12-August-2017]}</note>
</bibl>

<bibl id="B35">
  <title><p>A Survey on Location Estimation Techniques for Events Detected in
  Twitter</p></title>
  <aug>
    <au><snm>Ozdikis</snm><fnm>O</fnm></au>
    <au><snm>O\u{g}uzt\"{u}z\"{u}n</snm><fnm>H</fnm></au>
    <au><snm>Karagoz</snm><fnm>P</fnm></au>
  </aug>
  <source>Knowl. Inf. Syst.</source>
  <publisher>New York, NY, USA: Springer-Verlag New York, Inc.</publisher>
  <pubdate>2017</pubdate>
  <volume>52</volume>
  <issue>2</issue>
  <fpage>291</fpage>
  <lpage>-339</lpage>
  <url>https://doi.org/10.1007/s10115-016-1007-z</url>
</bibl>

<bibl id="B36">
  <title><p>A Survey of Location Inference Techniques on Twitter</p></title>
  <aug>
    <au><snm>Ajao</snm><fnm>O</fnm></au>
    <au><snm>Hong</snm><fnm>J</fnm></au>
    <au><snm>Liu</snm><fnm>W</fnm></au>
  </aug>
  <source>J. Inf. Sci.</source>
  <publisher>Thousand Oaks, CA, USA: Sage Publications, Inc.</publisher>
  <pubdate>2015</pubdate>
  <volume>41</volume>
  <issue>6</issue>
  <fpage>855</fpage>
  <lpage>-864</lpage>
  <url>http://dx.doi.org/10.1177/0165551515602847</url>
</bibl>

<bibl id="B37">
  <title><p>Geotagging Text Content With Language Models and Feature
  Mining</p></title>
  <aug>
    <au><snm>Kordopatis Zilos</snm><fnm>G</fnm></au>
    <au><snm>Papadopoulos</snm><fnm>S</fnm></au>
    <au><snm>Kompatsiaris</snm><fnm>I</fnm></au>
  </aug>
  <source>Proceedings of the {IEEE}</source>
  <publisher>Institute of Electrical and Electronics Engineers
  ({IEEE})</publisher>
  <pubdate>2017</pubdate>
  <volume>105</volume>
  <issue>10</issue>
  <fpage>1971</fpage>
  <lpage>-1986</lpage>
  <url>https://doi.org/10.1109/jproc.2017.2688799</url>
</bibl>

<bibl id="B38">
  <title><p>Web-a-where: Geotagging Web Content</p></title>
  <aug>
    <au><snm>Amitay</snm><fnm>E</fnm></au>
    <au><snm>Har'El</snm><fnm>N</fnm></au>
    <au><snm>Sivan</snm><fnm>R</fnm></au>
    <au><snm>Soffer</snm><fnm>A</fnm></au>
  </aug>
  <source>Proceedings of the 27th Annual International ACM SIGIR Conference on
  Research and Development in Information Retrieval</source>
  <publisher>New York, NY, USA: ACM</publisher>
  <series><title><p>SIGIR '04</p></title></series>
  <pubdate>2004</pubdate>
  <fpage>273</fpage>
  <lpage>-280</lpage>
  <url>http://doi.acm.org/10.1145/1008992.1009040</url>
</bibl>

<bibl id="B39">
  <title><p>Placing Flickr Photos on a Map</p></title>
  <aug>
    <au><snm>Serdyukov</snm><fnm>P</fnm></au>
    <au><snm>Murdock</snm><fnm>V</fnm></au>
    <au><snm>Zwol</snm><fnm>R</fnm></au>
  </aug>
  <source>Proceedings of the 32Nd International ACM SIGIR Conference on
  Research and Development in Information Retrieval</source>
  <publisher>New York, NY, USA: ACM</publisher>
  <series><title><p>SIGIR '09</p></title></series>
  <pubdate>2009</pubdate>
  <fpage>484</fpage>
  <lpage>-491</lpage>
  <url>http://doi.acm.org/10.1145/1571941.1572025</url>
</bibl>

<bibl id="B40">
  <title><p>You Are Where You Tweet: A Content-based Approach to Geo-locating
  Twitter Users</p></title>
  <aug>
    <au><snm>Cheng</snm><fnm>Z</fnm></au>
    <au><snm>Caverlee</snm><fnm>J</fnm></au>
    <au><snm>Lee</snm><fnm>K</fnm></au>
  </aug>
  <source>Proceedings of the 19th ACM International Conference on Information
  and Knowledge Management</source>
  <publisher>New York, NY, USA: ACM</publisher>
  <series><title><p>CIKM '10</p></title></series>
  <pubdate>2010</pubdate>
  <fpage>759</fpage>
  <lpage>-768</lpage>
  <url>http://doi.acm.org/10.1145/1871437.1871535</url>
</bibl>

<bibl id="B41">
  <title><p>Inferring Twitter User Locations with 10 Km Accuracy</p></title>
  <aug>
    <au><snm>Ryoo</snm><fnm>K</fnm></au>
    <au><snm>Moon</snm><fnm>S</fnm></au>
  </aug>
  <source>Proceedings of the 23rd International Conference on World Wide
  Web</source>
  <publisher>New York, NY, USA: ACM</publisher>
  <series><title><p>WWW '14 Companion</p></title></series>
  <pubdate>2014</pubdate>
  <fpage>643</fpage>
  <lpage>-648</lpage>
  <url>http://doi.acm.org/10.1145/2567948.2579236</url>
</bibl>

<bibl id="B42">
  <title><p>Find Me if You Can: Improving Geographical Prediction with Social
  and Spatial Proximity</p></title>
  <aug>
    <au><snm>Backstrom</snm><fnm>L</fnm></au>
    <au><snm>Sun</snm><fnm>E</fnm></au>
    <au><snm>Marlow</snm><fnm>C</fnm></au>
  </aug>
  <source>Proceedings of the 19th International Conference on World Wide
  Web</source>
  <publisher>New York, NY, USA: ACM</publisher>
  <series><title><p>WWW '10</p></title></series>
  <pubdate>2010</pubdate>
  <fpage>61</fpage>
  <lpage>-70</lpage>
  <url>http://doi.acm.org/10.1145/1772690.1772698</url>
</bibl>

<bibl id="B43">
  <title><p>SPOT: Locating Social Media Users Based on Social Network
  Context</p></title>
  <aug>
    <au><snm>Kong</snm><fnm>L</fnm></au>
    <au><snm>Liu</snm><fnm>Z</fnm></au>
    <au><snm>Huang</snm><fnm>Y</fnm></au>
  </aug>
  <source>Proc. VLDB Endow.</source>
  <publisher>VLDB Endowment</publisher>
  <pubdate>2014</pubdate>
  <volume>7</volume>
  <issue>13</issue>
  <fpage>1681</fpage>
  <lpage>-1684</lpage>
  <url>http://dx.doi.org/10.14778/2733004.2733060</url>
</bibl>

<bibl id="B44">
  <title><p>Tweets from Justin Bieber's Heart: The Dynamics of the Location
  Field in User Profiles</p></title>
  <aug>
    <au><snm>Hecht</snm><fnm>B</fnm></au>
    <au><snm>Hong</snm><fnm>L</fnm></au>
    <au><snm>Suh</snm><fnm>B</fnm></au>
    <au><snm>Chi</snm><fnm>EH</fnm></au>
  </aug>
  <source>Proceedings of the SIGCHI Conference on Human Factors in Computing
  Systems</source>
  <publisher>New York, NY, USA: ACM</publisher>
  <series><title><p>CHI '11</p></title></series>
  <pubdate>2011</pubdate>
  <fpage>237</fpage>
  <lpage>-246</lpage>
  <url>http://doi.acm.org/10.1145/1978942.1978976</url>
</bibl>

<bibl id="B45">
  <title><p>Mapas de las Provincias y Regiones de Chile. {Cartograf\'ia
  Rulamahue}</p></title>
  <aug>
    <au><snm>Albers</snm><fnm>C</fnm></au>
  </aug>
  <pubdate>2016</pubdate>
  <note>Available from: {\url{http://www.rulamahue.cl/}} [Accessed
  18-October-2017]</note>
</bibl>

<bibl id="B46">
  <title><p>Comparing Media Systems: Three Models of Media and
  Politics</p></title>
  <aug>
    <au><snm>Hallin</snm><fnm>DC</fnm></au>
    <au><snm>Mancini</snm><fnm>P</fnm></au>
  </aug>
  <publisher>Cambridge University Press</publisher>
  <series><title><p>Communication, Society and Politics</p></title></series>
  <pubdate>2004</pubdate>
  <url>https://books.google.it/books?id=954NJChZAGoC</url>
</bibl>

<bibl id="B47">
  <title><p>Random Forests</p></title>
  <aug>
    <au><snm>Breiman</snm><fnm>L</fnm></au>
  </aug>
  <source>Mach. Learn.</source>
  <publisher>Hingham, MA, USA: Kluwer Academic Publishers</publisher>
  <pubdate>2001</pubdate>
  <volume>45</volume>
  <issue>1</issue>
  <fpage>5</fpage>
  <lpage>-32</lpage>
  <url>https://doi.org/10.1023/A:1010933404324</url>
</bibl>

<bibl id="B48">
  <title><p>Lo Prado. {Wikipedia, The Free Encyclopedia}</p></title>
  <aug>
    <au><cnm>Wikipedia</cnm></au>
  </aug>
  <pubdate>2017</pubdate>
  <note>Available from: {\url{https://es.wikipedia.org/wiki/Lo\_Prado}
  [Accessed 12-August-2017]}</note>
</bibl>

<bibl id="B49">
  <title><p>San Miguel (Chile). {Wikipedia, The Free Encyclopedia}</p></title>
  <aug>
    <au><cnm>Wikipedia</cnm></au>
  </aug>
  <pubdate>2017</pubdate>
  <note>Available from:
  {\url{https://es.wikipedia.org/wiki/San\_Miguel\_(Chile)} [Accessed
  12-August-2017]}</note>
</bibl>

<bibl id="B50">
  <title><p>ICVU 2017: \'Indice de Calidad de Vida Urbana. Comunas y Ciudades
  de Chile</p></title>
  <aug>
    <au><snm>Orellana</snm><fnm>A</fnm></au>
  </aug>
  <publisher>Santiago, Chile</publisher>
  <pubdate>2017</pubdate>
  <note>Available from:
  {\url{http://fadeu.uc.cl/images/noticias/2017/05.Mayo/Presentacion_ICVU_2017_.pdf}}
  [Accessed 10-January-2018]</note>
</bibl>

<bibl id="B51">
  <title><p>Chile Descentralizado y Desarrollado: Fundamentos y Propuestas para
  Construir una Pol\'itica de Estado en Descentralizaci\'on y Desarrollo
  Territorial en Chile.</p></title>
  <aug>
    <au><snm>Baer</snm><fnm>H</fnm></au>
    <au><snm>Torralbo</snm><fnm>F</fnm></au>
  </aug>
  <source>95 propuestas para un Chile mejor</source>
  <publisher>Temuco, Chile: Instituto de Desarrollo Local y Regional, IDER.
  Universidad de La Frontera</publisher>
  <pubdate>2013</pubdate>
  <note>Available from:
  {\url{http://95propuestas.cl/site/wp-content/uploads/2013/05/chile-descentralizado-y-desarrollado-heinrich-von-baer-y-felipe-torralbo.pdf}}
  [Accessed 10-January-2018]</note>
</bibl>

<bibl id="B52">
  <title><p>Cuentas Nacionales de Chile. {PIB} Regional 2016</p></title>
  <aug>
    <au><snm>Chile</snm><fnm>BC</fnm></au>
  </aug>
  <pubdate>2017</pubdate>
  <note>Available from:
  \url{http://www.bcentral.cl/documents/20143/32019/CCNNPIB_Regional2016.pdf}
  [Accessed 30-December-2017]</note>
</bibl>

<bibl id="B53">
  <title><p>Radio Cooperativa (Chile). {Wikipedia, The Free
  Encyclopedia}</p></title>
  <aug>
    <au><cnm>Wikipedia</cnm></au>
  </aug>
  <pubdate>2017</pubdate>
  <note>Available from:
  {\url{https://es.wikipedia.org/wiki/Radio\_Cooperativa\_(Chile)} [Accessed
  30-December-2017]}</note>
</bibl>

<bibl id="B54">
  <title><p>Ranking General de Audiencia Gran Santiago. Ipsos Radio
  2016.</p></title>
  <aug>
    <au><cnm>{Ipsos}</cnm></au>
  </aug>
  <pubdate>2015</pubdate>
  <note>Available from: {\url{http://www.ipsos.cl/ipsosradioalaire/pagdos.htm}}
  [Accessed 30-December-2017]</note>
</bibl>

<bibl id="B55">
  <title><p>El Quinto Poder. {FD+D}</p></title>
  <aug>
    <au><cnm>{Fundaci\'on Democracia y Desarrollo}</cnm></au>
  </aug>
  <pubdate>2010</pubdate>
  <note>Available from: {\url{http://www.elquintopoder.cl/} [Accessed
  30-December-2017]}</note>
</bibl>

<bibl id="B56">
  <title><p>Media Mergers and Media Bias with Rational Consumers</p></title>
  <aug>
    <au><snm>Anderson</snm><fnm>SP</fnm></au>
    <au><snm>McLaren</snm><fnm>J</fnm></au>
  </aug>
  <source>Journal of the European Economic Association</source>
  <publisher>Blackwell Publishing Inc</publisher>
  <pubdate>2012</pubdate>
  <volume>10</volume>
  <issue>4</issue>
  <fpage>831</fpage>
  <lpage>-859</lpage>
  <url>http://dx.doi.org/10.1111/j.1542-4774.2012.01069.x</url>
</bibl>

<bibl id="B57">
  <title><p>Handcuffs for the Grabbing Hand? Media Capture and Government
  Accountability</p></title>
  <aug>
    <au><snm>Besley</snm><fnm>T</fnm></au>
    <au><snm>Prat</snm><fnm>A</fnm></au>
  </aug>
  <source>American Economic Review</source>
  <pubdate>2006</pubdate>
  <volume>96</volume>
  <issue>3</issue>
  <fpage>720</fpage>
  <lpage>736</lpage>
  <url>http://www.aeaweb.org/articles?id=10.1257/aer.96.3.720</url>
</bibl>

<bibl id="B58">
  <title><p>Detection of User Demographics on Social Media: A Review of Methods
  and Recommendations for Best Practices</p></title>
  <aug>
    <au><snm>Cesare</snm><fnm>N</fnm></au>
    <au><snm>Grant</snm><fnm>C</fnm></au>
    <au><snm>Nsoesie</snm><fnm>EO</fnm></au>
  </aug>
  <source>CoRR</source>
  <pubdate>2017</pubdate>
  <volume>abs/1702.01807</volume>
  <url>https://arxiv.org/abs/1702.01807</url>
</bibl>

</refgrp>
} 




\section*{Figures}

\begin{figure}
\centering
    \begin{subfigure}[b]{0.475\textwidth}
        \centering
        \caption[]%
        {Population}
        \label{fig:population}
    \end{subfigure}
    \hfill
    \begin{subfigure}[b]{0.475\textwidth}
        \centering
        \caption[]%
        {Average distance to news outlets}
        \label{fig:z-dist}
    \end{subfigure}
    \vskip\baselineskip
    \begin{subfigure}[b]{0.475\textwidth}
        \centering
        \caption[]%
        {Income}
        \label{fig:z-income}
    \end{subfigure}
    \quad
    \begin{subfigure}[b]{0.475\textwidth}
        \centering
        \caption[]%
        {Right-leaning}
        \label{fig:z-right-leaning}
    \end{subfigure}
    \caption[]
    {\csentence{Summary of four different metrics for Chile}. From left to right, top to bottom: population, distance, income and political-leaning}
    \label{fig:maps}
\end{figure}

\begin{figure}
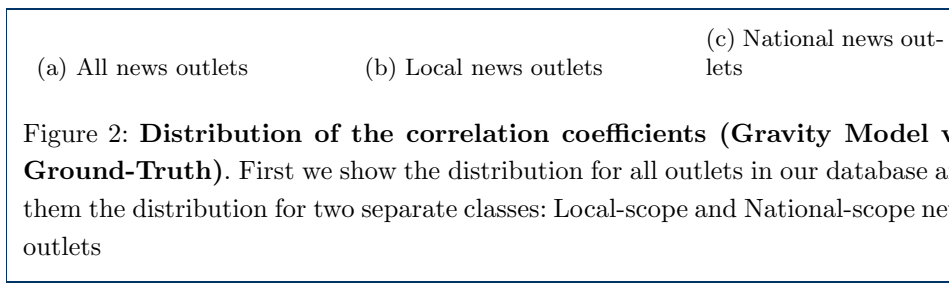

\centering
\begin{subfigure}[b]{0.25\textwidth}
    \centering
    \caption{All news outlets} \label{fig:gm_tw_corr_all}
\end{subfigure}
\hfill
\begin{subfigure}[b]{0.25\textwidth}
    \centering
    \caption{Local news outlets} \label{fig:gm_tw_corr_local}
\end{subfigure}
\hfill
\begin{subfigure}[b]{0.25\textwidth}
    \centering
    \caption{National news outlets} \label{fig:gm_tw_corr_national}
\end{subfigure}
\caption{\csentence{Distribution of the correlation coefficients (Gravity Model vs. Ground-Truth)}. First we show the distribution for all outlets in our database and them the distribution for two separate classes: Local-scope and National-scope news outlets}
\label{fig:gm_tw_corr}
\end{figure}

\begin{figure}
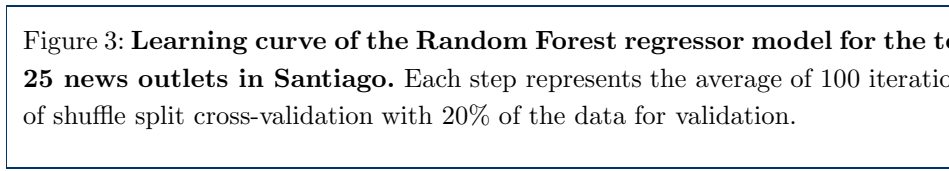

    \centering
    \caption{\csentence{Learning curve of the Random Forest regressor model for the top 25 news outlets in Santiago.} Each step represents the average of 100 iterations of shuffle split cross-validation with 20\% of the data for validation.} \label{fig:learning_curve}
\end{figure}

\begin{figure}
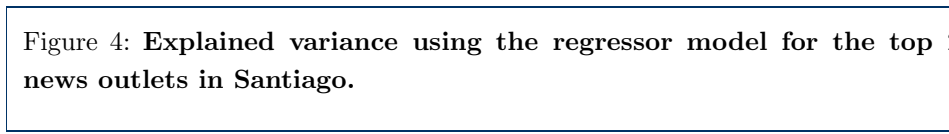

    \centering
    \caption{\csentence{Explained variance using the regressor model for the top 25 news outlets in Santiago.}} 
    \label{fig:exp_variance_25}
\end{figure}

\begin{figure}
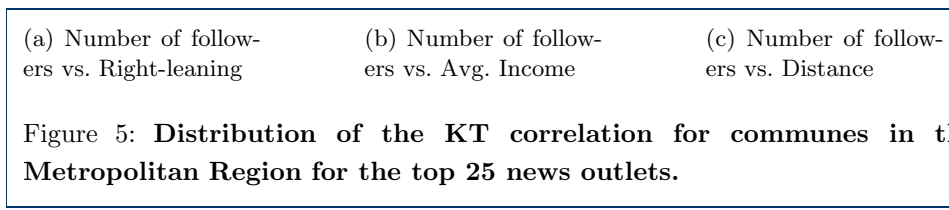

\centering
\begin{subfigure}[b]{0.25\textwidth}
    \centering
    \caption{Number of followers vs. Right-leaning} \label{fig:kt_corr_political}
\end{subfigure}
\hfill
\begin{subfigure}[b]{0.25\textwidth}
    \centering
    \caption{Number of followers vs. Avg. Income} \label{fig:kt_corr_economic}
\end{subfigure}
\hfill
\begin{subfigure}[b]{0.25\textwidth}
    \centering
    \caption{Number of followers vs. Distance} \label{fig:kt_corr_geo}
\end{subfigure}
\caption{\csentence{Distribution of the KT correlation for communes in the Metropolitan Region for the top 25 news outlets.}}
\label{fig:kt_corr}
\end{figure}

\begin{figure}
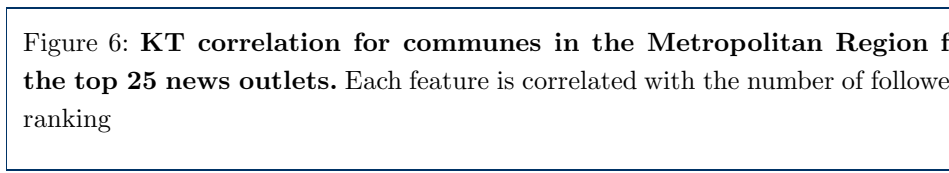

    \centering
    \caption{\csentence{KT correlation for communes in the Metropolitan Region for the top 25 news outlets.} Each feature is correlated with the number of followers' ranking} \label{fig:kt_all_followers}
\end{figure}

\begin{figure}
    \centering
    \caption{\csentence{KT correlation for communes in the Metropolitan Region for top 25 \textit{newspapers}.} Each feature is correlated with the number of followers ranking} \label{fig:kt_all_followers_nps}
\end{figure}

\begin{figure}
    \centering
    \caption{\csentence{KT correlation for communes in the Metropolitan Region comparing News Outlets and Football players behaviour.} Each feature is correlated with the number of followers' ranking of the corresponding dataset.} \label{fig:kt_players_vs_outles}
\end{figure}





\section*{Tables}

\begin{table}[h!]
\caption{Summary of news outlets and football players followers on Twitter.}
\label{table:followers}
      \begin{tabular}{lrr}
        \hline
                                            & \# Outlets' Followers & \# Players' Followers \\ \hline
        Unique users                        & 4,943,351  (100\%)    & 6,568,769  (100\%)    \\
        Users w/ non-empty location         & 1,579,068  (31\%)     & 2,434,183  (37\%)     \\
        Users w/ useful location            & 996,326    (20\%)     & 540,828    (8\%)      \\ 
        Users w/ GPS coord                  & 4,829      (0.1\%)    & 2,041      (0.03\%)   \\ 
        Users w/ high confidence location   & 597,981    (12\%)     & 383,207    (6\%)     \\ \hline
      \end{tabular}
\end{table}

\begin{table}[h!]
\caption{Stats about the news outlets' correlation coefficients.}
\label{table:stats}
\begin{tabular}{lrrr}
    \hline
     Category   & Total &  $\rho >0.7$ & $\rho <0.2$ \\
    \hline
     Outlets &  402 & 203 & 141\\
     w/ nacional scope &  133 & 23 & 94\\
     w/ local scope &  269 & 180 & 47\\
     located in Santiago &  156 & 29 & 108\\
     w/ nacional scope \& located in Santiago &  126 & 18 & 93\\
    \hline
\end{tabular}
\end{table}

\begin{table}[h!]
\caption{Correlation between features}
\label{table:corr}
\begin{tabular}{lrrr}
    \hline
     Feature   & \tt{right-leaning} & \tt{income} & \tt{distance} \\
    \hline
     \tt{right-leaning}  &  1.00 &  0.59 & -0.48\\
     \tt{income}         &  0.59 &  1.00 & -0.35\\
     \tt{distance}       & -0.48 & -0.35 &  1.00\\
    \hline
\end{tabular}
\end{table}




\end{backmatter}
\end{document}